\documentclass[review,onefignum,onetabnum]{siamonline171218}

\usepackage[mathscr]{euscript}
\usepackage{amsmath,amssymb}

\usepackage[algo2e,lined,boxed,commentsnumbered]{algorithm2e}
\usepackage{subfigure}
\usepackage{soul}
\usepackage{bigstrut}
\usepackage{multirow}
\usepackage{multicol}
\usepackage{booktabs}
\usepackage[normalem]{ulem}
\newcommand{\stkout}[1]{\ifmmode\text{\sout{\ensuremath{#1}}}\else\sout{#1}\fi}

\DeclareMathOperator{\EX}{\mathbb{E}}
\DeclareMathOperator*{\argmax}{argmax}
\DeclareMathOperator*{\argmin}{argmin}

\usepackage{lipsum}
\usepackage{amsfonts}
\usepackage{graphicx}
\usepackage{epstopdf}
\usepackage{algorithmic}
\ifpdf
  \DeclareGraphicsExtensions{.eps,.pdf,.png,.jpg}
\else
  \DeclareGraphicsExtensions{.eps}
\fi

\usepackage{enumitem}
\setlist[enumerate]{leftmargin=.5in}
\setlist[itemize]{leftmargin=.5in}

\newsiamremark{remark}{Remark}
\newsiamremark{hypothesis}{Hypothesis}
\crefname{hypothesis}{Hypothesis}{Hypotheses}
\newsiamthm{claim}{Claim}

\headers{DQN for Optimal Execution}{B. Ning, F. Lin and S. Jaimungal}

\title {
Double Deep Q-Learning for Optimal Execution\thanks{SJ would like to acknowledge the support of the Natural Sciences and Engineering Research Council of Canada (NSERC), [funding reference numbers RGPIN-2018-05705 and RGPAS-2018-522715]}
}

\author{Brian Ning \thanks{Department of Statistical Sciences, University of Toronto
  (\email{brian.ning@mail.utoronto.ca}, \email{sebastian.jaimungal@utoronto.ca}).}
\and Franco Ho Ting Lin \thanks{Department of Computer Science, University of Toronto
  (\email{francohtlin@cs.toronto.edu}).}
\and Sebastian Jaimungal \footnotemark[1]}

\usepackage{amsopn}

\makeatletter
\newcommand*{\addFileDependency}[1]{
  \typeout{(#1)}
  \@addtofilelist{#1}
  \IfFileExists{#1}{}{\typeout{No file #1.}}
}
\makeatother

\newcommand*{\vcenteredhbox}[1]{\begingroup
\setbox0=\hbox{#1}\parbox{\wd0}{\box0}\endgroup}

\ifpdf
\hypersetup{
  pdftitle={DQN for Optimal Execution},
  pdfauthor={B. Ning, F. Lin and S. Jaimungal}
}
\fi
\nolinenumbers

\begin{document}

\maketitle

\begin{abstract}

Optimal trade execution is an important problem faced by essentially all traders. Much research into optimal execution uses stringent model assumptions and applies continuous time stochastic control to solve them. Here, we instead take a model free approach and develop a variation of Deep Q-Learning to estimate the optimal actions of a trader. The model is a fully connected Neural Network trained using Experience Replay and Double DQN with input features given by the current state of the limit order book, other trading signals, and available execution actions, while the output is the Q-value function estimating the future rewards under an arbitrary action. We apply our model to nine different stocks and find that it outperforms the standard benchmark approach on most stocks using the measures of (i) mean and median out-performance, (ii) probability of out-performance, and (iii) gain-loss ratios.

\end{abstract}



\section{Introduction}

Financial markets are highly complex stochastic systems with significant heteroskedasticity.
The problem of how to optimally execute large positions over a given trading horizon  is an important problem faced by institutional investors, banks, and hedge funds. Na\"ively rebalancing  a portfolio could result in significant adverse price movements as other intelligent traders may read off the signal. Investors must balance trading quickly and obtaining poor execution prices, with trading slowly which exposes them to unknown market fluctuations.

Agents are often exposed to a plethora of information including prior stock prices and other market conditions. Determining the best trading policy in the presence of all of this information is an important part of the  algorithmic trading literature. Traditionally, researchers propose a stochastic model based on empirical observations, such as an Ornstein-Uhlenbeck or stochastic volatility process, use historical data to estimate model parameters, such as the volatility, mean-reversion level and rate, propose a performance criteria which they aim to maximize, and then solve the problem analytically using methods in stochastic optimal control. One of the earliest works in this vein is the  Almgren-Chriss \cite{almgren2001optimal} approach, where the authors assume prices are Brownian motion. There has been many generalizations including of this approach to account for a variety of market features, see, e.g., \cite{gueant2015general}, \cite{cartea2016incorporating}, and \cite{casgrain2018}, and the graduate textbook \cite{cartea2015algorithmic} for a modern treatment.

However, in the case of more complex price models without analytical solutions, or even non-parametric models, a different approach is necessary. Reinforcement learning \cite{Sutton:1998}  attempts to learn optimal policies for sequential decision problems by optimizing a cumulative future reward function with few modeling assumptions (such as Markovian structure of the state space). The most popular reinforcement learning algorithm is Q-learning \cite{Watkins1992} which has been applied to the optimal execution problem in \cite{Nevmyvaka:2006} and \cite{rl_ac}.

One of the major drawbacks to the Q-learning algorithm is that it  estimates the optimal policy on a pointwise basis. This severely limits the ability to interpolate between possible actions, and therefore does not allow for effective generalization. Moreover, the algorithm's storage space scales exponentially with respect to the dimension of the state space. To address this issue, we modify recent advancements in Deep Q-Learning \cite{DBLP:journals/corr/MnihKSGAWR13,van2016deep} to make it useful for the solving the optimal execution problem. By estimating the Q-function used in Q-learning with a deep neural network, we can interpolate between actions and states as well as reduce the storage costs associated with classical Q-learning.

To the best of our knowledge this article is the first to adapt and modify the framework of Deep Q-learning to the optimal execution problem. We also  provide numerical comparisons between our approach and classical methods, as well as discuss the financial interpretation of the results.

The remainder of this paper is structured as follows. Section 2 provides some background information on key concepts including a brief description of Q-learning and the optimal execution problem. Section 3 and 4 details the exact formulation of the optimal execution problem in a reinforcement learning setting and the adaption of Deep Q-learning. Section 5 explains how we train the network with a detailed algorithm. Section 6 and 7 presents our results on real data and the metrics we use to evaluate our findings.

\section{Background}

\subsection{Limit Order Books}

\begin{figure}[h]
	\centering
\subfigure[Before MO Arrival]{\includegraphics[width=0.3\textwidth]{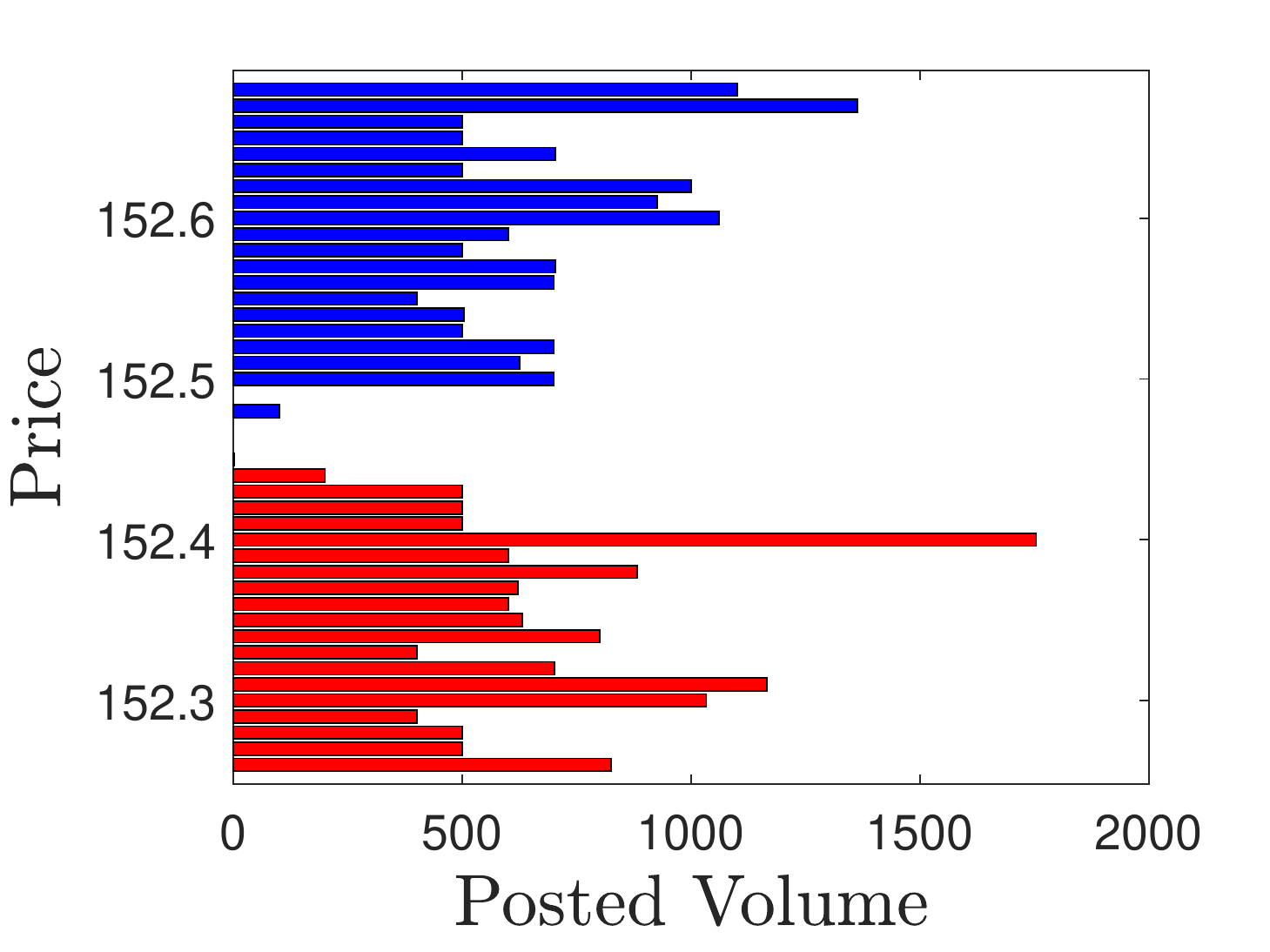}}
\subfigure[Buy MO walking the LOB]{\includegraphics[width=0.3\textwidth]{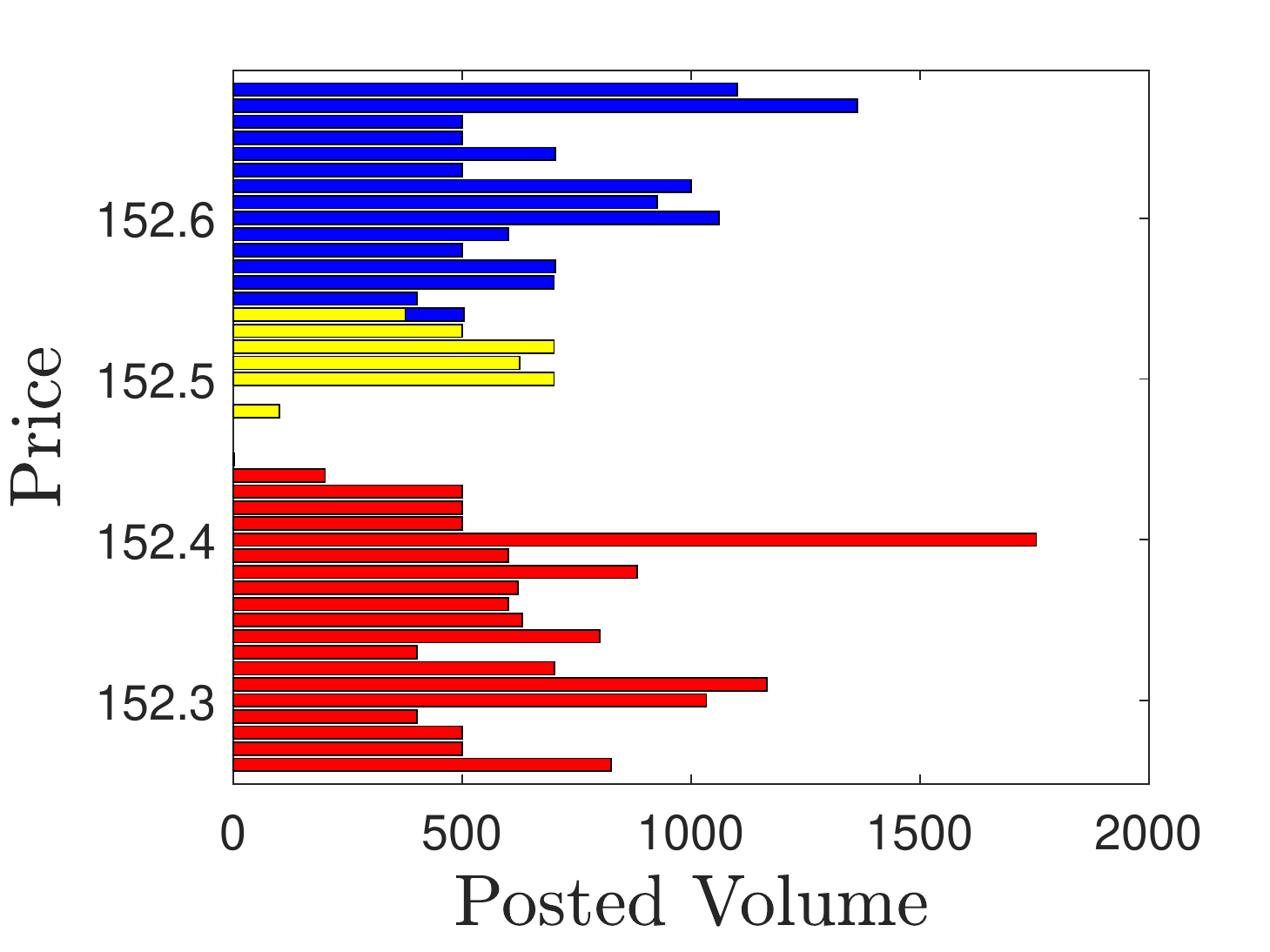}}
\subfigure[After MO Arrival]{\includegraphics[width=0.3\textwidth]{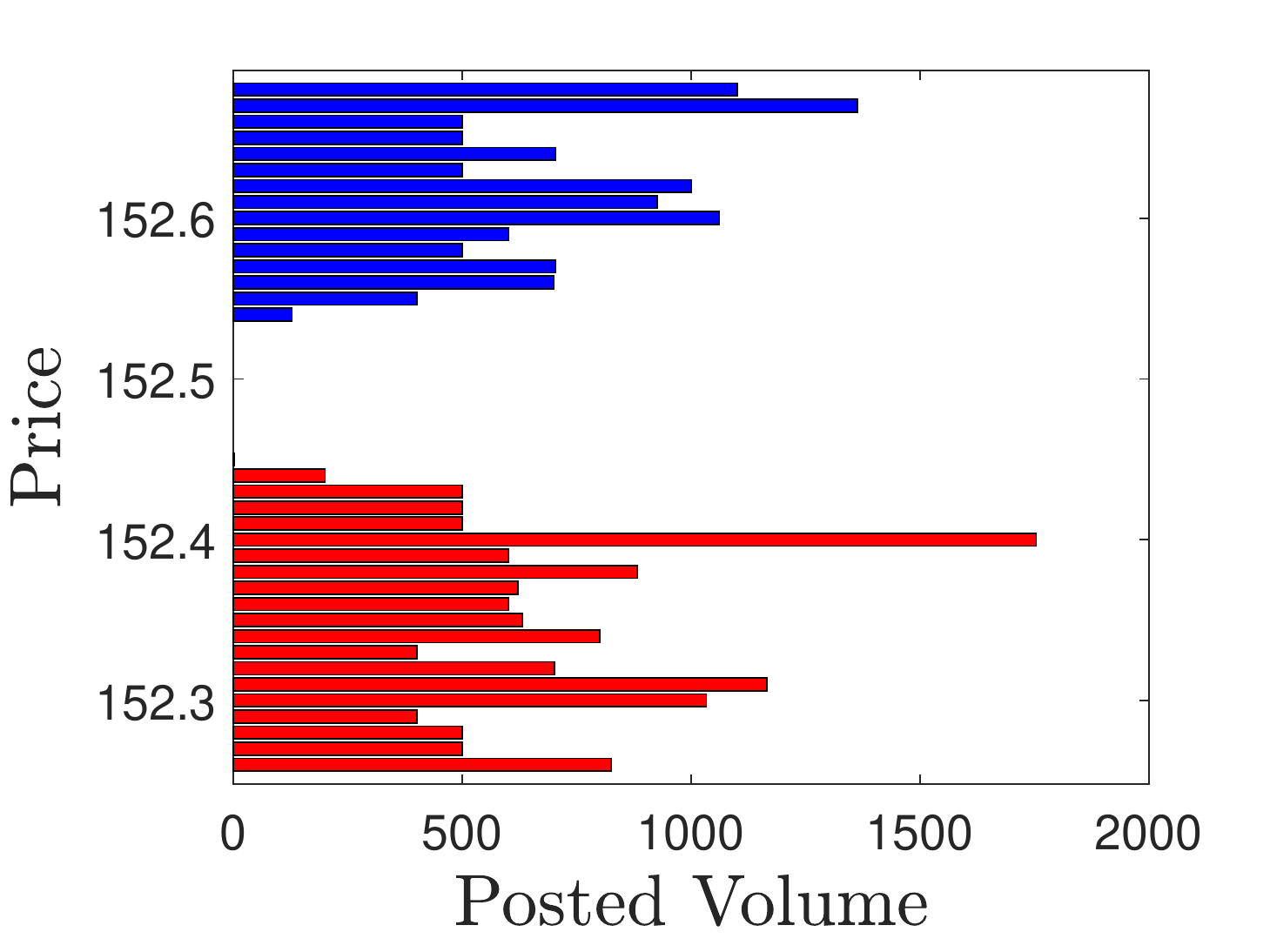}}
	\caption{The LOB of Facebook Inc. stock at 10:36 on 28 Mar, 2018 as a buy MO of size 3,000 arrives. The blue (red) bars represent available volume of sell (buy) orders at the corresponding price. The yellow bars in the middle panel indicate which LOs match the incoming MOs.}
	\label{fig:LOB-Snapshot}
\end{figure}
Traders may execute two main order types  on modern electronic stock exchanges: \textit{limit orders} and \textit{market orders}. A limit order(LO) represents the intention to buy or sell at most a fixed number of shares at a specified price. If there are no shares available at that specified price (or better), then the limit order is collected in the \textit{limit order book} (LOB) and  remains there until the trader either cancels the order or it is filled by another trader. Therefore LOs have a price guarantee, but are not guaranteed to be executed. Traders who post limit orders are said to provide liquidity to the market as they provide a pool of shares at different prices for other traders to either buy or sell. A market order(MO) represents a fixed number of shares to buy or sell, and is matched by the best available LOs remaining in the LOB. The  lowest price of all sell LOs is known as the \textit{ask price} and the highest price of all buy LOs is known as the \textit{bid price}. The difference between the ask and bid is known as the \textit{spread} and the average of the ask and bid price is known as the \textit{mid-price}. The minimum change in the bid or ask is called the \textit{tick size}. Figure {fig:LOB-Snapshot} illustrate a snapshot of the LOB for INTC as a buy MO walks the LOB. Before the MO arrives, the LOB has a spread of 1 tick, the bid and ask are \$48.40 and \$48.41, respectively. After the MO walks the LOB, the bid and ask are \$48.40 and \$48.46, respectively, and the spread widens to 6 ticks.

The share prices in our data are ``small'' relatively to the tick size, and hence the spread is typically 1 tick -- such stocks are called \textit{large tick stocks}. Therefore, we approximate all execution prices by the mid-price and ignore the spread. We avoid the cost associated with walking the LOB by applying a penalty on the size of any one order, thus the trader's strategy only takes the liquidity posted at the best available price. The details are described in the paragraph before \eqref{eqn:one-step-reward}.

\subsection{Optimal Execution}

The optimal execution problem assumes studies a trader who at  $t= 0$ holds an inventory of $q_0$ shares and must  fully liquidated it by the end of the time period $t = T$ (i.e. $q_T = 0$). We study the problem in discrete times, so that trade orders $x_t \geq 0$ are executed at evenly distributed time-steps $t = \{0, 1, 2, ... , T-1\}$ where $x_t$ represents the number of shares to be sold. The midprice $p_t$ follows some unknown process which may or may not be impacted by the trader's actions. In our case, however, we  assume  the trader's actions do not directly effect the price process during training. This is a reasonable assumption to make as the volume of shares traded for most stocks of large companies heavily exceeds that of individual investors. However once the network is trained on historical data and used to perform online learning in real-time, it can learn what effect our trading action has on the price process and adapt the optimal policy as needed. We define $s_t = [p_t, \overline{s}_t]^\intercal$ where $\overline{s}_t$ is a vector of stochastic processes representing other features. Price and other states are jointly  stochastic and may be affected by the actions taken at time $t$ so that $s_{t+1}=f(S_t,x_t)$ is random. The goal is to maximize the expected total profit obtained by selling the shares subject to transactions fees
\begin{equation}
\argmax_{x_0,x_1,...,x_{t-1}} \EX \left[ \sum_{t=0}^{T-1} R(p_t,x_t)\right]\,,
\end{equation}
where $R(s_t,x_t)$ is the profit obtained from selling $x_t$ shares at mid-price $p_t$. Note, the only feature which feeds into $R$  is   the midprice. We  work with the full state vector $s_t$ from this point onwards and  $\overline{s}$ are various selected features as discussed in Section \ref{sec:features}.

\subsection{Reinforcement Learning}

The goal in reinforcement learning is to learn a policy $\pi \in \Pi$ to control a system with states $s \in \mathscr{S}$ using actions $x\in \mathscr{X}$ in order to maximize the total expected rewards based on some reward function $R(s,x)$. The system itself is defined by an initial state distribution $P(s_0)$ and transition distribution $P(s_t|s_{t-1},x_{t-1})$. The goal is to maximize the total expected discounted rewards defined as $R = \sum_{t=1}^T \gamma ^{t-1} R(s_t,x_t)$ over the space of all allowable policies $\Pi$. As optimal execution problems are often performed over short time periods, the discount factor is set close to one.

\subsection{Deep Q-Learning}
In Q-learning, the Q-function \cite{Watkins1992}, given an optimal policy $\pi \in \Pi$, is defined as:
\begin{equation}
 Q(s,x) = \EX\left[R(s,x) + \sum_{i=t+1}^T \gamma ^{i-t} R(s_i^{\pi},x_i^{\pi})\right]\,,
\end{equation}
where the action at time $t$ is arbitrary, but from $t+1$ onwards it is optimal with $s_i^{\pi}$ and $x_i^{\pi}$ representing the state and action respectively at time $i$ if the agent follows the optimal policy $\pi$. Under mild conditions, the Q-function satisfies the Bellman equation
\begin{equation}
 Q(s,x) = \EX\left[R(s,x) +  \gamma \max_{x'\in\mathcal{U}} Q(s^{s,x},x')\right],
\end{equation}
where $s^{s,x}$ is the (random) state the system evolves to after taking action $x$ when in state $s$, and $\mathcal{U}$ is the admissible set of actions. It may be estimated in an online iterative fashion as follows   (see \cite{Sutton:1998})
\begin{equation}
\label{eqn:Q-update-tradational}
Q^{(\ell+1)}(s,x) = Q^{(\ell)}(s,x) + \alpha_t\left[ R(s,x) + \gamma\, \max_{x' \in \mathcal{U}}Q^{(\ell)}(s^{s,x},x') - Q^{(\ell)}(s,x) \right]\,,
\end{equation}
where $Q^{(\ell)}(s,x)$ denotes the estimate of $Q^{(\ell)}(s,x)$ at iteration $\ell$,  $R(s,x)$ is a realized reward by taking action $x$ in state $s$, and $s^{s,x}$ is the (random) state the system evolves to after taking the action,
as long as $\sum_{t=0}^\infty \alpha_t =\infty$ and $\sum_{t=0}^\infty \alpha_t^2 <+\infty$.

When $\mathscr{S}$ and $\mathscr{X}$ are discrete and low dimensional, it is possible to represent the Q-function as a matrix. For larger dimensions or continuum, however, it is typically replaced with a model approximation. In Deep Q-Learning, this model is a fully connected neural network $Q(s,x|\theta)$, where $\theta$ are the network parameters rather than using \eqref{eqn:Q-update-tradational} to update the network.

At each iteration of the algorithm, the network parameters are updated by minimizing the squared loss between the Q-value using the previous network parameters $\theta_\ell$ and the Q-value using updated network parameters. The loss function
\begin{equation}
L(\theta;\theta_\ell) = \left(
\left[
R(s,x) + \gamma \max_{x' \in \mathcal{U}}Q\left(\left.s^{s,x},x'\,\right|\,\theta_{\ell}\right)
\right]
- Q\left(\left.s,x\,\right|\,\theta\right)\right)^2
\label{equation:loss}
\end{equation}
is minimized at each iteration and the new network parameters $\theta_{\ell+1} = \argmin_{\theta} L(\theta;;\theta_\ell)$.

\section{Optimal Execution in a Reinforcement Learning Setting}
\label{sec:optimal_execution}

In this work, we focus on demonstrating how to use our approach when the trader employs  market orders only. An interesting follow up would be to incorporate an optimal mix of limit and market orders.  The restriction to market orders is sub-optimal, as market orders incur a cost due to the bid-ask spread, as well as a cost due to walking the limit order book (as in Figure \ref{fig:LOB-Snapshot}). Nonetheless, even with this restriction, we  demonstrate that our approach provides gains over time weighted average price (TWAP) -- which is optimal under the assumption that price is a Brownian motion, and more generally a martingale.



In the next subsections, we provide a description of the states, actions and rewards used in our reinforcement learning formulation of the optimal execution problem. The execution time horizon is divided into \textit{T} periods and  execution decisions are made at the beginning of each period. Trades, however, are made each second.

Specifically, we denote the time periods where trading decisions can be made by $T_0<T_1<T_2< \dots<T_{N-1}$. We also denote the end of the last trading period as $T_N = T$, however, the last decision is made at time $T_{N-1}$. The intra-period time, where trades occur, are indexed by $\{\{t_{0,i}\}_{i\in\{0,\dots,M_0-1\}},\dots,\{t_{N-1,i}\}_{i\in\{0,\dots,M_{N-1}-1\}}\}$, with  $t_{k,i} = T_{k} + i\Delta t$, where $\Delta t=\frac{T_k-T_{k-1}}{M_k}$ so that period $k$ has trading time indices $T_{k}=t_{k,0}<t_{k,1}<\dots<t_{k,M_k-1}=T_{k+1}-\Delta t$ and is made up of $M_k$ small time steps. In the numerical experiments we have $T_k-T_{k-1} = c, \, \forall \, k$ for some constant $c$ and $\Delta t = 1\,sec$.



\subsection{States}  The state space (or feature space) $\mathscr{S}$ contains the state of the LOB at the start of each period, as well as any prior information from previous periods. We make the (standard) assumption that given a state $s_t\in\mathscr{S}$ and a trading action $x_t\in\mathscr{X}$ at time $t$, the mapping $(s_t,x_t) \mapsto s_{t+1}$ is Markovian. Time inhomogeneous processes may be incorporated by allowing one of the components of the state space to be  time $t$ itself. We explore a variety of features  in the results section. The two states which play a crucial role in the optimal strategy are (i) the current time (or elapsed time) $t$ and (ii) the remaining inventory $q_t$ to execute, and they are always part of our  state space.


\subsection{Actions} A policy uniquely maps a state $s\in\mathscr{A}$ to an action $x\in\mathscr{X}$. In particular, in the optimal execution problem, the only action is the amount of shares to sell via a market order. Naturally, the set of allowed actions  is restricted such that $x_t \in[0, q_t]$ for all $t$. We make a further restriction on the action space to take only integer values (or fixed multiples of integer values for larger inventories) in order to simplify the computation needed to find the optimal action at each time step.
At each executable time step for actions, the total amount of shares sold are assumed to be evenly distributed over the time block on a second by second basis.

\subsection{Rewards} The reward  equals the total reward over each period, and each period reward is made of rewards for trading each second. Over each period, we assume the agent sends orders at a constant rate, so that when the trader makes a decision at time $T_k$ to send $x_{T_k}$ orders over the next period, the actual trades are made of $\frac{x_{T_k}}{M_k}$ equal trades every second. To account for transaction costs and the potential that these shares may walk the LOB (as in Figure \ref{fig:LOB-Snapshot}), we also incorporate a quadratic penalty on the number of shares executed during each second. This penalty term can adjusted with a hyperparameter to represent the liquidity of the asset in question. One may also replace it with any other non-linear penalty that reflects the agents utility associated with taking on impact risk. Specifically, for each intra-period timestep $[t_{k,i}, t_{k,i+1})$ the reward is
\begin{equation}
\check{R}_{k,i} = q_{t_{k,i}} \left(p_{t_{k,i+1}}-p_{t_{k,i}}\right) - a \left(\tfrac{x_{T_k}}{M_k}\right)^2\,,
\label{eqn:one-step-reward}
\end{equation}
where $q_{t_{k,i}}$ and $p_{t_{k,i}}$ are the remaining inventory and price at time $t_{k,i}$, respectively. The total reward over the period $[T_k,T_{k+1})$ is the sum of each intra-period reward,
\begin{equation}
R_k = \sum_{i=0}^{M_k-1} \check{R}_{k,i}\,,
\label{eqn:PeriodReward}
\end{equation}
and the total reward  is the sum of each period reward
\begin{align}
\sum_{k=0}^{N-1} R_k = \sum_{k=0}^{N-1} \sum_{i=0}^{M_k-1} \check{R}_{k,i}
&=\sum_{k=0}^{N-1} \sum_{i=0}^{M_k-1} \left\{q_{t_{k,i}} (p_{t_{k,i+1}}-p_{t_{k,i}}) - a \left(\frac{x_{T_k}}{M_k}\right)^2\right\}\\
&=-q_{0}p_{0} + \sum_{k=0}^{N-1} \sum_{i=0}^{M_k-1} \left\{ p_{t_{k,i+1}} \frac{x_{T_k}}{M_k} - a \left(\frac{x_{T_k}}{M_k}\right)^2\right\}
\label{eqn:sum-one-step-reward}
\end{align}
where the last equality follows as $q_T=0$,  $q_t-q_{t-1} = x_t$, and by using  a summation by parts formula. The total reward is  the amount sold at each time period less a  penalty based on the amount sold.

\section{Network Architecture}

\begin{figure}
  \centering
  \includegraphics[scale=0.45]{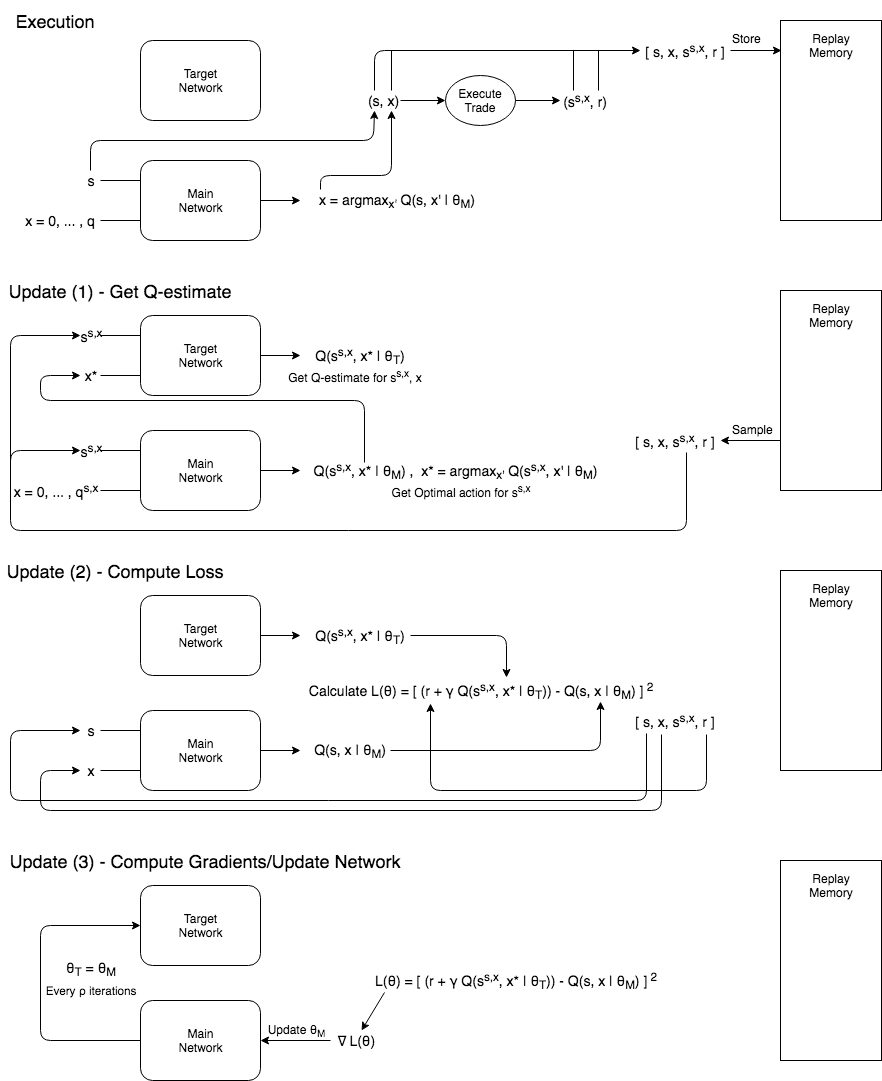}
  \caption{Execution procedure and full update procedure used when training the network. Note that when we are training our network, $\epsilon$-Greedy is added to the Execution step. In the update procedure, we split the process into 3 steps, (1) Getting the Q-estimate given the next state $s^{s,x}$, (2) Computing the Loss after we have obtained the Q-estimate and (3) Computing the gradients and updating our network weights.}
  \label{fig:qnet}
\end{figure}

In this section we describe our  network architecture and the update procedure we use to train the networks. Figure \ref{fig:qnet} illustrates the various steps in the training procedure and how the training and target networks used in our  double deep Q-learning framework relate to one another.

\subsection{Basic Network Architecture}
\label{sec:basicnet}

As actions in the optimal execution problem are (in principle) continuous, our network architecture uses both states and action as inputs and the Q-value function as output. If instead, we discretise or bucket actions, it would be possible to use states only as inputs, and have a network for each action; however, as the allowed actions depends on the state (recall that $x_t\in[0,q_t]$) making this approach less tractable.  Another possible architecture is to have multiple Q-value function as outputs, with each corresponding to the specific action taken. Since we have varying action spaces at each time step, the logical approach is to use a network that takes both a state and action as input.


The network architecture has dimension $s + 1$, where $s\in\{2,3,4,5\}$ corresponds to the number of features and the extra dimension is the action. The collection of features we use are: time, inventory, price, quadratic variation, and price. They are discussed further in Section \ref{sec:Experiments}. As the input dimension is low, we  use a fully connected feed-forward neural network with 6 layers and 20 nodes in each layer. At each layer, we use a ReLU activation function to prevent vanishing gradients and provide sparsity, and we use the RMSprop optimizer (see \cite{DBLP:journals/corr/MnihKSGAWR13}). 

\subsection{Experience Replay}
\label{sec:expreplay}
In order to improve network stability in the case of extreme values in recently visited states, a replay memory is used to provide a database of previously visited combinations of state and action to sample from in order to perform the network weight updates. \cite{Lin:1992} Specifically, at each time step the tuple $d = (s,x,r,s^{s,x})$ is stored in the replay memory buffer $\mathscr{D}$. Where $s^{s,x}$ denotes the state we arrive at from $s$ when action $x$ is taken. Batches are then sampled uniformly from $\mathscr{D}$ and used to compute the gradient of the loss function in \eqref{equation:loss} in order to update the network. Once the replay memory has reached a certain size $K$, we randomly remove a transition from the first $\frac{K}{2}$ tuples. This is done to ensure that we do not remove cases that were immediately placed into the buffer which may not have been selected by random sampling to update the network yet.

\subsection{Main and Target Networks}
\label{sec:doubleNet}

As noted in \cite{van2016deep}, the original form of DQN may suffer from oscillations in the gradient updates, as the same network $Q(s,x|\theta)$ generates both the next state's target Q-values and updates the current state's Q-values. To reduce this instability, we use two neural networks: a main network $Q(s,x\,|\,\theta_M)$ and a target network $Q(s,x\,|\,\vartheta_T)$. We update the main network at the end of each decision period $t=T_{k+1}$, $k \in {0, 1, \dots , N-1}$, while the target network $\vartheta_T$ remains fixed and is replaced by $\theta_M$ only after several periods of trading. As well, we compute gradients  from samples in our replay memory $\mathscr{D}$ instead of the immediate transitions observed. Thus the loss function is written as
\begin{equation}
\label{eqn:maintargetloss}
L(\theta; \vartheta_T) = \sum_{j = 1} ^ J \left(
\left[
r_{(j)} + \gamma \max_{x' \in [0, q_{j}-x_{j}]}
Q\left(\left.s_{(j)}^{s,x},x'\,\right|\,\vartheta_T\right)
\right] - Q\left(\left.s_{(j)},x_{(j)}\,\right|\,\theta\right)\right)^2,
\end{equation}
and $\theta_M = \argmin_{\theta} L(\theta;\vartheta_T)$
where $(s_{(j)},x_{(j)},r_{(j)},s_{(j)}^{s,x})$ is sampled from the memory replay $\mathscr{D}$ for all $j = 1\dots J$ described in \ref{sec:expreplay}.

\subsection{Double DQN}
Main-target networks often overestimate the Q-values and we correct this by using Double DQN \cite{van2016deep}. We decouple the first term from our loss function in \eqref{eqn:maintargetloss} into action selection and action evaluation by using our main network $Q(s,x|\theta_M)$ to select the best action, and our target network $Q(s,x|\vartheta_T)$ to generate our Q-value estimate. This is in contrast to  using the target network for both selecting the best action and generating the Q-value. Our modified estimate of the Q function is
\begin{equation}
    Q\left(\left. s_{(j)}^{s,x}, \;x^*\!\!\left(\left.s_{(j)}^{s,x}\,\right|\,\theta_M\right)\right| \vartheta_T\right),
    \qquad
    \text{where}
    \qquad
    x^*\!\!\left(\left.s_{(j)}^{s,x}\,\right|\,\theta_M\right)= \displaystyle{\argmax_{x'\in[0,q_{(j)}]}} Q\left(\left.s_{(j)}^{s,x},x'\,\right|\,\theta_M\right),
\end{equation}
our modified loss function is
\begin{equation}
\label{eqn:maintargetloss-mod}
L(\theta; \vartheta_T) = \sum_{j = 1} ^ J \left(
\left[
r_{(j)} + \gamma\, Q\left(\left. s_{(j)}^{s,x}, x^*\!\!\left(\left.s_{(j)}^{s,x}\,\right|\,\theta_M\right) \right| \vartheta_T\right)
\right] - Q\left(\left.s_{(j)},x_{(j)}\,\right|\,\theta\right)\right)^2,
\end{equation}
the updated main network minimizes the loss function $\theta_M = \argmin_{\theta} L(\theta;\vartheta_T)$, and $(s_{(j)}$, $x_{(j)}$, $r_{(j)}$, $s_{(j)}^{s,x})$ is sampled from the memory replay $\mathscr{D}$ for all $j = 1\dots J$.

\subsection {Treatment of the Zero Ending Inventory Constraint}
\label{sec:zeroending}
In DQN, the network structure estimates the Q-function at all time periods, including the last period. In the optimal execution problem, however, the last period has the additional constraint that inventory must be drawn down to zero by the end of the period. In classical Q-Learning, with a standard matrix representation of the Q-function, such constraints are imposed analytically and the terminal period reward determines the terminal Q-function
\begin{equation}
 Q(s_{T_{N-1}},x_{T_{N-1}}) = R_{N-1}\,. 
\end{equation}

However, as discussed in Section \ref{sec:optimal_execution} each period is made of  one-second intervals, and an action taken at time $T_k$ results in trades at each second $\{t_{k,i}\}_{i=0,\dots,M_k-1}$ in that period. At the start of the last period $T_{N-1}$, the reward $R_{N-1}$ in Equation (\ref{eqn:PeriodReward}) depends on the price path $\{s_{t_{N-1,i}}\}_{i \in 0,\dots ,M_{N-1}-1}$ which are not measurable (except for $i=0$) at time $T_{N-1}$ when the action is to be taken. Thus there is no deterministic value for the last timestep as the reward at this timestep depends on how the price process behaves in the intra-period intervals from the action-executable time to the end of the period.

There are several ways to  address this issue. One approach is to assume  a model for the intra-period price process $s_{t_{N-1,i}} \sim \pi_0\left(\cdot\left|s_{t_{N-1,i-1}},\,\frac{q_{T_{N-1}}}{M_{N-1}}\right.\right)$ during the last period, which enforces the constraint $q_{T_N}=0$ and implies that $x_{T_{N-1}}= q_{T_{N-1}}$, and therefore
\begin{equation}
Q(s_{T_{N-1}},x_{T_{N-1}}) =\EX \left[\left.\sum_{i = 0}^{M_{N-1}-1} \check{R}_{N-1,i}^{\pi_0,x_{T_{N-1}}}\, \right|\,s_{T_{N-1}}\right],
\end{equation}
where $\check{R}_{N-1,i}^{\pi_0,x_{T_{N-1}}}$ is the intra-period reward (as defined in \eqref{eqn:one-step-reward}) obtained when the state evolves according to the law of $\pi_0$ after executing action $x_{T_{N-1}}$ at time $T_{N-1}$.

A second approach is to estimate the Q-value of this last interval using the neural network $ Q(s_{T-1},x_{T-1}) \approx  Q(s_{T-1},x_{T-1}|\theta)$ and  enforce the restriction that the only action allowable at the last action-executable time  equals  the remaining inventory.

We take an alternate approach that places less burden on the network to estimate terminal Q-values. We add a single one second time step at the end of the trading horizon over which all remaining shares are liquidated. To incorporate this terminal liquidation into the learning procedure, whenever a selected state from the reply buffer corresponds to the last time period, we replace the correspond term in the loss in  Equation \eqref{eqn:maintargetloss} with
\begin{equation}
 \left(
\left[
r_{(j)} + \gamma \, \mathfrak{R}(s_{(j)}^{s,x},q_{(j)}-x_{(j)}) + \gamma\, Q\left(\left. s_{(j)}^{s,x}, x^*\!\!\left(\left.s_{(j)}^{s,x}\,\right|\,\theta_M\right) \right| \vartheta_T\right)
\right] - Q\left(\left.s_{(j)},x_{(j)}\,\right|\,\theta\right)\right)^2
\end{equation}
where the terminal reward (from liquidating all remaining shares) is
\begin{equation}
\mathfrak{R}(s,q) =  q \, (p'-p) - a\,q^2\,,
\end{equation}
and $p'$ is the price at time $T+\Delta T$ reached from the state $s_{(j)}^{s,x}$ at time $T_{N}$. This specific loss function is only applied to the experiences sampled from the replay memory where the initial state corresponds to time $t = T_{N-1}$. All other sampled experiences uses the original loss function defined in \eqref{eqn:maintargetloss}. The total batch loss  the sum of the individual losses of each experience sampled.
This approach makes no assumption on the price process, unlike the first approach. It  also places much less burden on the network to correctly estimate  terminal Q-values, as would be the case using the second approach. One shortcoming is that the constraint of liquidating all shares by $T_N$ is not strictly enforced, although it is enforced by $T_N+\Delta T$. For any reasonable value of $a$ (the coefficient of the quadratic penalty term on size of execution in the rewards function), however, the optimal action results in selling all (or near all) remaining inventory and thus satisfying the restriction in the original problem.


\section{Training Method}
In this section, we specify how we train the network as outlined in Algorithm \ref{alg:trainingprocess}.
\begin{algorithm2e}[h!]
\caption{Double DQN Optimal Execution Training Method}
\label{alg:trainingprocess}
Initialize replay memory $\mathscr{D}$ of size $\mathcal{N}$\;
Initialize action-value function $Q(\cdot|\theta_M)$ with random weights.
Pre-train $Q(\cdot|\theta_M)$ on boundary cases and make a copy $Q(\cdot|\vartheta_T)$\;
\For{trading episode $b \in B$}
{
    \For{$i\leftarrow 0$ \KwTo $N-1$ }
    {
        With probability $\epsilon$ select random action $x_i \sim \text{Binomial}\left(q_i,\frac{1}{T_i}\right)$\;
        Otherwise select $x_i = \displaystyle{\max_{x'\in[0,q_{T_i}]}}Q(s_{T_i},x'\,|\,\theta_M)$ (optimal action)\;
        Execute the action $x_{T_i}$ and observe the reward $r_{T_i}$ and next state $s_{{T_{i+1}}}^{s_{T_i},x_{T_i}}$\;
        Store transition $(s_{T_i},x_{T_i},r_{T_i},s_{T_{i+1}}^{s_{T_i},x_{T_i}})$ in replay buffer $\mathscr{D}$\;
        \For{$j\leftarrow 1$ \KwTo $J$}
        {
            Sample random minibatch of transitions $\left(s_{(j)},x_{(j)},r_{(j)},s_{(j)}^{s,x}\right)$ from $\mathscr{D}$\;
            \BlankLine
            \begin{eqnarray*}
            \text{Set } y_{(j)}(\vartheta_T) =
            \begin{cases}
            r_{(j)}, & \mbox{for $t_{(j)}^{s,x} = T_{N}$}
            \\
            r_{(j)} + \gamma \, \mathfrak{R}(s_{(j)}^{s,x},q_{(j)}), & \mbox{for $t_{(j)}^{s,x} = T_{N-1}$}
            \\
            r_{(j)} + \gamma\,Q\left(\left.s_{(j)}^{s,x}, x^*_{(j)} \right| \vartheta_T\right), &\mbox{otherwise}
             \end{cases}
            \end{eqnarray*}
            where $q_{(j)}$ is the inventory remaining corresponding to the state $s_{(j)}$ and $t_{(j)}^{s,x}$ is the time remaining corresponding to the state $s_{(j)}^{s,x}$
            \[
            x_{(j)}^*=\displaystyle{\argmax_{x'\in[0,q_{(j)}]}} \; Q\left(\left.s_{(j)}^{s,x},x'\,\right|\,\theta_M\right)\;;
            \]
        }
        Obtain new network parameters $\theta_M$ by minimizing
        \[
        L(\theta; \vartheta_T) = {\sum_{j = 1} ^ J} \left[y_{(j)}(\vartheta_T) - Q\left(\left.s_{(j)},x_{(j)}\,\right|\,\theta\right)\right]^2
        \]
        using gradient descent to obtain $\theta_M = \argmin_{\theta} L(\theta;\vartheta_T)$\;
    }
    Decay $\epsilon = \tau \epsilon$\;
    After $\rho$ iterations update $\theta_M = \vartheta_T$\;
}
\end{algorithm2e}

\subsection{$\epsilon$-Greedy}
Reinforcement Learning require trading off exploration versus exploitation. \cite{Sutton:1998} Exploration allows the system to evolve into regions in state space which have not yet been sampled. These regions may have larger rewards than what the model would otherwise tell us. Once state space is sufficiently explored, we can be more certain what actions are better than others, and we can exploit this knowledge by greedily selecting the best action given the state we are in.

For the exploration/exploitation trade off we use $\epsilon$-greedy \cite{Sutton:1998} actions. Specifically, we set $\epsilon\in(0,1)$, let $\xi \sim Unif(0,1)$, and use the policy
\begin{eqnarray}
x_{T_k} = \begin{cases} \text{Binomial}\left(q_{T_k},\frac{1}{T_N-T_k}\right)\,, &
\mbox{if } \xi < \epsilon
\\[0.5em]
 \displaystyle{\max_{x\in[0,q_{T_k}]}}Q\left(\left.s_{T_k},x\,\right|\,\vartheta_T \right)\,, & \mbox{otherwise.}
\end{cases}
\label{eq:egreedy}
\end{eqnarray}
This selects the current estimate of the optimal action an average of $(1-\epsilon)$ of the time and a random action sampled from a Binomial the other $\epsilon$ of the time. We decay $\epsilon \leftarrow \tau \epsilon$, with $\tau<1$, because as we observe more data, the network estimate of Q is more accurate, and we wish to exploit more often. The $\epsilon$-Greedy approach relies on random exploration as opposed to other distribution based approaches like Boltzmann Exploration or by using a Bayesian Neural Network.\cite{Sutton:1998}  For the $\epsilon$-Greedy action, we choose a binomial distribution so that (i) the action respects the constraint $x\in[0,q]$, (ii) on average, the selected action equals $\frac{q_{T_k}}{T_N-T_k}$ which is a TWAP strategy for the remaining inventory, and (iii) the action space is sufficiently explored.

\subsection{Hyperparameters}
As seen in Algorithm \ref{alg:trainingprocess}, there are a number of hyperparameters aside from the usual neural network hyperparameters. Additionally, there is (i) the replay memory size $\mathcal{N}$, (ii) the rate of decay of $\epsilon$ given as $\tau$, and (iii) the rate at which to update the target and main network $\rho$. These hyperparameters cannot be tuned with cross-validation due to the training procedure for reinforcement learning algorithms and are therefore carefully selected. Another important hyperparameter specific to the optimal execution problem is the quadratic penalty coefficient $a$. The parameter $a$ is selected based on the liquidity of the stock and the trader's estimated transaction costs.

\subsection{Pre-training}

In order to increase network stability during training, the network is first trained on a set of boundary action cases using randomly selected price intervals from the full data set. These boundary cases are: hold the full inventory then sell all shares at the last time step, and sell all shares at the at the first time step.

\section{Features}
\label{sec:features}

The state space is defined by a set of features we chose to observe from the market and used as inputs to our neural network. In the following section, we will explain each input variable and the transformations applied to project them into the range $[-1,1]$.

\begin{figure}[t!]
  \centering
  \includegraphics[scale=0.45]{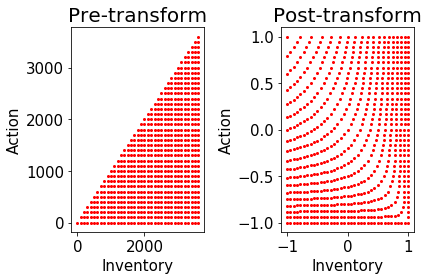}
  \caption{Inventory and Action Transformation. For explicit formula see Appendix \ref{sec:transform-q-x-domain}.}
  \label{fig:trans}
\end{figure}

The two key features are time and inventory. It is vital to track how much time remains to execute trades and how much inventory remains to be executed. We transform time by an affine transformation so that it lies in the interval $[-1,1]$. As the domain of allowable actions is restricted to be less than the inventory remaining at the start of the same period, the original domain of inventory/action pairs lie in a triangular region in inventory/action space (see the left panel of Figure \ref{fig:trans}). To increase stability of the algorithm, we transform the triangular region into the domain $\mathcal{K}=[-1,1]\times[-1,1]$.
The individual points may be seen in Figure \ref{fig:trans} and details of the transformation can be found in Appendix \ref{sec:transform-q-x-domain}.

Other important features include price, quadratic variation and price trend. Price $p_{t_{k,i}}$ is taken as the midprice at the end of each second. We subtract price at the beginning of each hour from the midprice, and perform an affine transformation so that only outliers in the historical prices lie outside the range $[-1,1]$. We denote this feature as the transformed price $(\tilde{P})$.    Quadratic Variation (QV) \cite{BarndorffNielsen2002} is a measure of the volatility of the asset prize and we anticipate that volatility affects the optimal strategy. To incorporate this effect, We use QV from the previous period as a feature, and estimate it as
\begin{equation}
QV_{T_k} = \sum_{i=0}^{M_{k-1}-1}\left(p_{t_{k-1,i}}-p_{t_{k-1,i-1}}\right)^2,\qquad \forall\, k\in\{1,\dots,N-1\}
\end{equation}
The initial QV value, $QV_{T_0}$, is defined using price data from the period before trading begins. To ensure the feature corresponding to QV lies in the interval $[-1,1]$, we transform the QV by  subtracting the mean and scaling by twice the standard deviation.

\section{Results}
\label{sec:Results}

\subsection{Data}

We test our approach on data using all trading days from 2-January-2017 to 30-March-2018 and stocks  Apple (AAPL), Amazon (AMZN), Facebook (FB), Google (GOOG), Intel (INTC), Microsoft (MSFT), NetApp Inc. (NTAP), Market Vectors Semiconductor ETF (SMH) and Vodophone (VOD). We use the full limit order book information to extract the midprice at the end of each second. As well, we exclude the most volatile times of the day and are able to avoid the diurnal patterns by analyzing the hours  11am-12pm, 12pm-1pm, 1pm-2pm of each day separately. These trading hours are referred to as hours 11, 12 and 13.


\subsection{Evaluation Metrics}
To evaluate the performance of our solution, we use Profit and Loss with Transaction Cost (P\&L)
computed for each trading hour $b \in \mathcal{B}=\{11,12,13\}$  as follows
\begin{equation}
\begin{aligned}
    {P\&L}_b &= \sum_{k=0}^{N-1} \sum_{i=0}^{M_k-1} \left\{ x_{t_{k,i}} p_{t_{k,i}} - a \left(\tfrac{x_{T_k}}{M_k}\right)^2\right\},
\end{aligned}
\end{equation}
where $\frac{x_{T_k}}{M_k}$ is the number of shares executed, $p_{t_{k,i}}$ the price at the start of the second, and the term $- a (\frac{x_{T_k}}{M_k})^2$ accounts for the penalty when executing large trades.

P\&L is computed for each hour of trading and compared to a Time-Weighted Average Price (TWAP) strategy (i.e. selling the same number of shares at each action-executable timestep). TWAP is often used a fair price of the asset of the duration of a trader, and is the resulting optimal strategy when agents believe the asset price process is a martingale \cite{almgren2001optimal}. As a measure of relative performance, we use the basis point improvement of P\&L relative to TWAP, defined as
\begin{equation}
    {\Delta P\&L}_{b} := \frac{{P\&L}_{b}^{\text{model}} - {P\&L}_{b}^{\text{TWAP}}}{{P\&L}_{b}^{\text{TWAP}}}\times 10^4, \qquad \forall b \in \mathcal{B}\,.
\label{equation:diffpnl}
\end{equation}
Given the relative performance for each day, and each trading hour, we report its mean, median, standard-deviation, gain-loss ratio \[
GLR:=\frac
{\mathbb{E}\left[\Delta P\&L\,|\, \Delta P\&L>0]\right]}
{\mathbb{E}\left[-\Delta P\&L\,|\, \Delta P\&L<0\right]}\,,
\] and the positive probability $\mathbb{P}(\Delta P\&L>0)$.

\subsection{Experiments}
\label{sec:Experiments}

For the experiments, we set $Q_0 = 20$ lots ($1$ lot equals $100$ shares) over a time horizon of $T = 1$ hour. We split the hour evenly into $N=5$ periods. At the beginning of each interval we compute the number of shares we wish to execute. The shares are then executed each second evenly across the entire interval. We test the neural network architecture specified in Section \ref{sec:basicnet} with a quadratic penalty $a=0.01$, main-target network update after $\rho = 15$ iterations, and a replay memory of size $\mathcal{N} = 10,000$. We fix our discount factor $\gamma=0.99$ for all tests.

\subsubsection{Time, Inventory}

The two key features that affect the rate of trading are Time and Inventory. Our results show that the optimal policy in this case is purely deterministic. This is consistent with the information filtration available to a trader who makes trades using only their current inventory and time. This is also consistent with the continuous time stochastic control approaches, see, e.g., \cite{almgren2001optimal}\cite{cartea2015algorithmic}.

\subsubsection{Time, Inventory, Price}
\label{sec:TIPresults}


\begin{figure}[t!]
  \centering
  \includegraphics[width=0.95\textwidth]{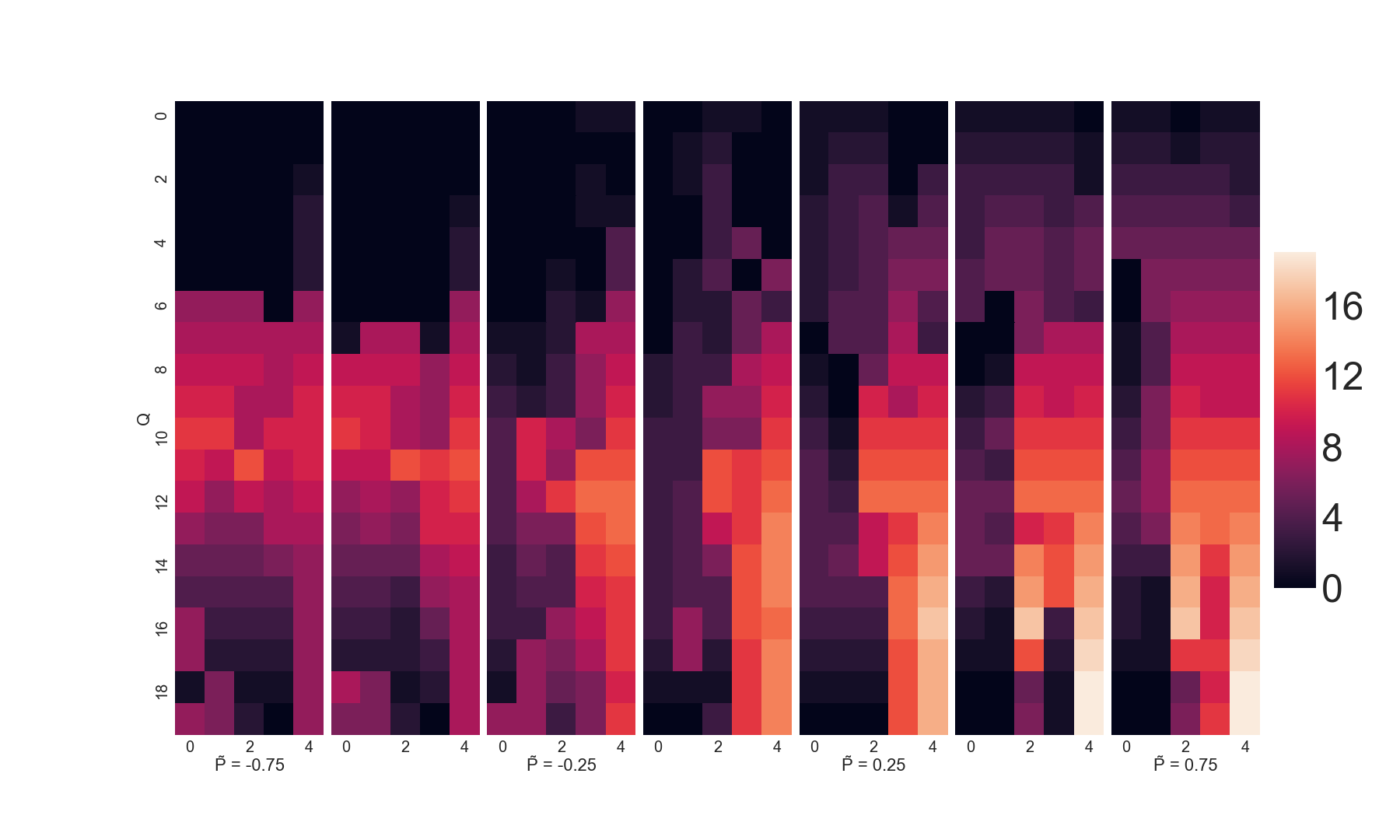}
  \vspace*{-2em}
  \caption{NTAP -- Optimal Actions for All Possible States with Features: Time, Inventory, Price --- Each plot from left to right denotes a change from low to high (normalized) price. Within each panel, the x and y axis denote time and inventory remaining, respectively. The color of individual squares denote the amount of shares sold, lighter being more shares, darker being less. }
  \label{fig:NTAP-TIP}
\end{figure}
Once a network model is trained on data, we iterate through all the time steps, inventory levels and price levels and compute the optimal actions traced out in every state. Those optimal actions are displayed in the heatmaps in Figure \ref{fig:NTAP-TIP}. As the figure shows, for the same level of inventory and time period, as we move from the left panel (lower prices) to the right panel (higher prices), the optimal strategy is to send more shares. As well, for any given panel, as we move from earlier times to later times, with the inventory remaining held fixed, the optimal strategy is to increase the shares executed. At the very last time period, all remaining shares are executed, regardless of inventory remaining or price. Finally, for a fixed time period, as the inventory remaining decreases, the optimal strategy is to execute less shares. All of these observed features of the optimal strategy are consistent with the trader aiming to execute all shares by the end of the time horizon, taking advantage of price improvements, all while managing their inventory risk.

\subsubsection{Time, Inventory, Price, Quadratic Variation}

Next, we add QV to assess how volatility alters the optimal solution.
\begin{figure}[h!]
  \centering
  \includegraphics[scale=0.35]{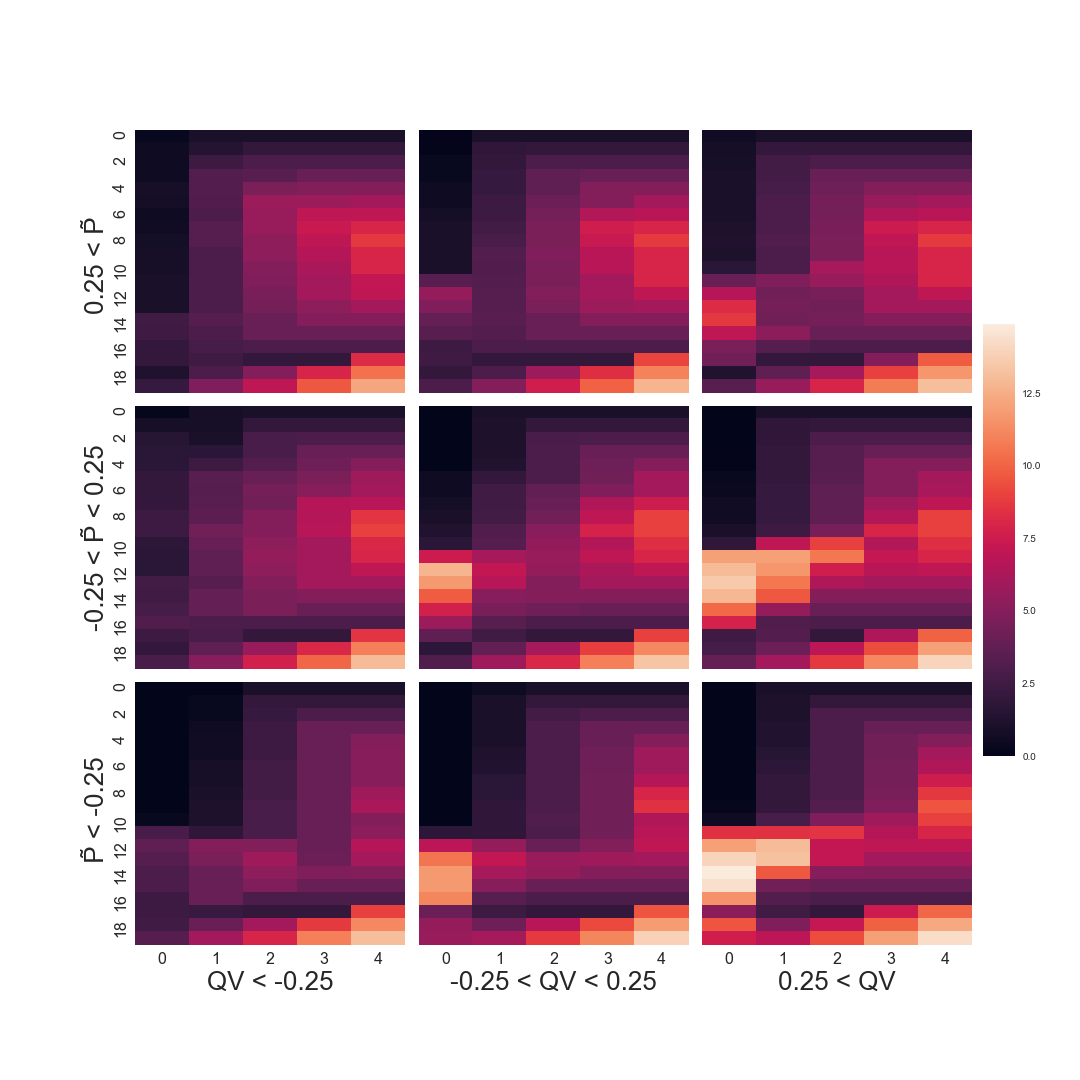}
  \caption{NTAP -- Average Optimal Actions Split on QV and Price with Features: Time, Inventory, Price, QV. --- For each individual plot, the x-axis denotes time intervals and y-axis inventory remaining. Moving across graphs, the x-axis denote changes from low to high QV, whereas the changes in the y-axis denote changes in price from low to high. }
  \label{fig:NTAP-TIPQV}
\end{figure}
To illustrate how the strategy is affected by time, price, inventory, and QV, in Figure \ref{fig:NTAP-TIPQV} we show heatmaps of the optimal strategy split across different price and QV levels, as a function of time and inventory. As QV increases, the number of executions increases regardless of the price. Periods of high volatility implies increased fluctuations in price, and therefore the agent wishes to get rid of their shares rather than be exposed to this risk. As prices move from low to high, there is a visible increase in the shares executed, which is consistent with the results in Section \ref{sec:TIPresults} and Figure \ref{fig:NTAP-TIP}.

\subsection{Statistical results}

Table \ref{tbl:all-results} shows various comparative statistics of the learned strategies using TIP and TIPQV for a collection of assets.
\begin{table}[h!]
\caption{Relative P\&L performance for all stocks with respect to TWAP strategy. \label{tbl:all-results}}
\begin{center}
    \begin{tabular}{clrrrrr}
    \toprule
    \toprule
          &       & \multicolumn{3}{c}{$\Delta P\&L$} &       &  \\
\cmidrule{3-5}    \multicolumn{1}{l}{Ticker} & Features & \multicolumn{1}{l}{Median} & \multicolumn{1}{l}{Mean} & \multicolumn{1}{l}{Std.Dev.} & \multicolumn{1}{l}{GLR} & \multicolumn{1}{l}{$\mathbb{P}(\Delta P\&L>0)$} \\
    \midrule
    \multirow{2}[2]{*}{AAPL} & TIP   & 2.92  & 2.85  & 6.1   & 1.0   & 77.4\% \\
          & TIPQV & 2.68  & 2.68  & 5.3   & 1.2   & 76.2\% \\
    \midrule
    \multirow{2}[2]{*}{AMZN} & TIP   & 0.06  & -0.28 & 11.3  & 0.9   & 50.1\% \\
          & TIPQV & -0.02 & -0.15 & 11.7  & 1.0   & 49.9\% \\
    \midrule
    \multirow{2}[2]{*}{FB} & TIP   & 2.61  & 2.52  & 7.6   & 1.2   & 68.1\% \\
          & TIPQV & 2.29  & 2.26  & 8.3   & 1.1   & 64.3\% \\
    \midrule
    \multirow{2}[2]{*}{GOOG} & TIP   & -0.59 & 0.21  & 7.4   & 1.3   & 45.7\% \\
          & TIPQV & -0.40 & 0.05  & 11.2  & 1.1   & 47.2\% \\
    \midrule
    \multirow{2}[2]{*}{INTC} & TIP   & 11.63 & 11.08 & 5.6   & 2.5   & 95.8\% \\
          & TIPQV & 11.96 & 11.40 & 4.0   & 3.6   & 97.8\% \\
    \midrule
    \multirow{2}[2]{*}{MSFT} & TIP   & 5.97  & 5.93  & 3.0   & 3.1   & 97.4\% \\
          & TIPQV & 5.89  & 5.95  & 3.8   & 2.8   & 94.8\% \\
    \midrule
    \multirow{2}[2]{*}{NTAP} & TIP   & 10.13 & 9.62  & 5.0   & 1.8   & 96.4\% \\
          & TIPQV & 10.05 & 9.64  & 5.0   & 2.1   & 96.2\% \\
    \midrule
    \multirow{2}[2]{*}{SMH} & TIP   & 3.80  & 2.67  & 6.7   & 1.0   & 73.5\% \\
          & TIPQV & 4.90  & 4.86  & 3.9   & 2.1   & 91.8\% \\
    \midrule
    \multirow{2}[2]{*}{VOD} & TIP   & 14.68 & 13.99 & 5.2   & 4.1   & 97.8\% \\
          & TIPQV & 15.72 & 15.43 & 3.1   & 18.6  & 99.2\% \\
    \bottomrule
    \bottomrule
    \end{tabular}%
\end{center}
\end{table}

While the results vary from one asset to the next, generally the relative P\&L has positive Mean and Median, as well, the gain-loss ratio is at least $1$, and  the probability of outperforming TWAP is high. Only Google and Amazon appear to have no significant improvement beyond TWAP. Moreover, adding quadratic variation generally improves the performance of the strategy relative to TWAP.

\begin{figure}[h!]
\setlength{\tabcolsep}{10pt}
\renewcommand{\arraystretch}{1.5}
\begin{center}
\begin{tabular}{rcc}
AAPL
&
\vcenteredhbox{\includegraphics[width=0.32\textwidth]{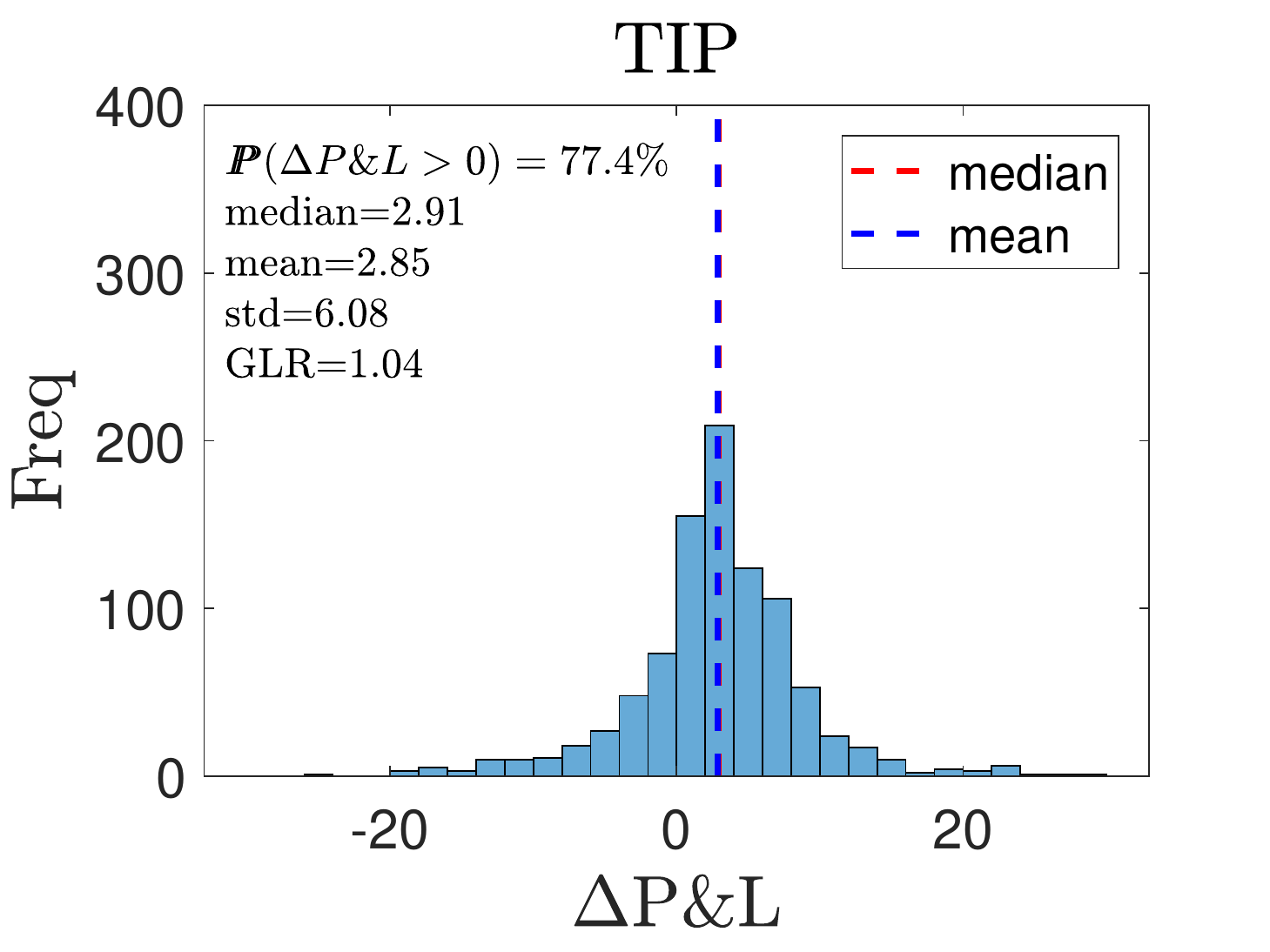}}
&
\vcenteredhbox{\includegraphics[width=0.32\textwidth]{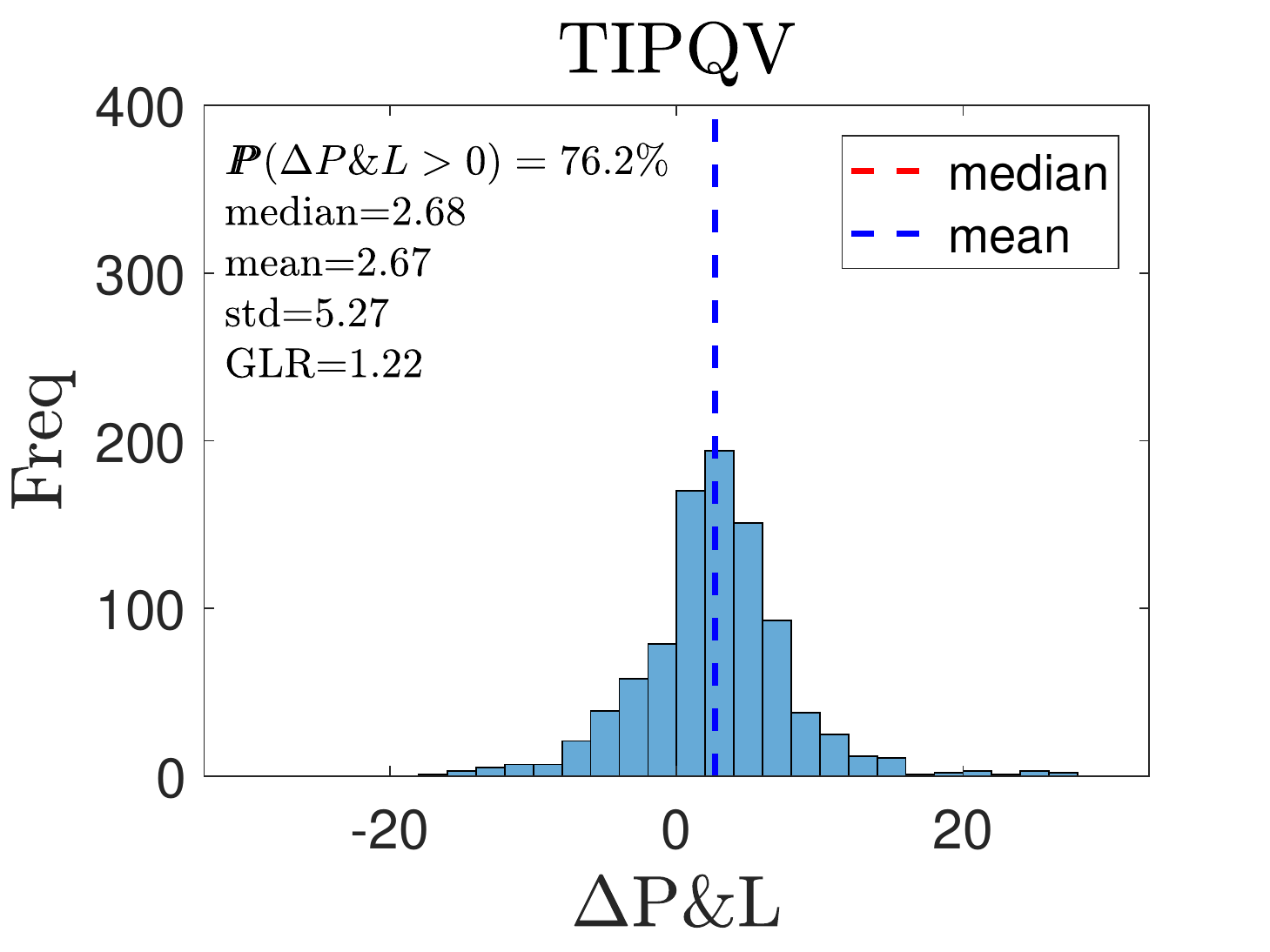}}
\\[1em]
AMZN &
\vcenteredhbox{\includegraphics[width=0.32\textwidth]{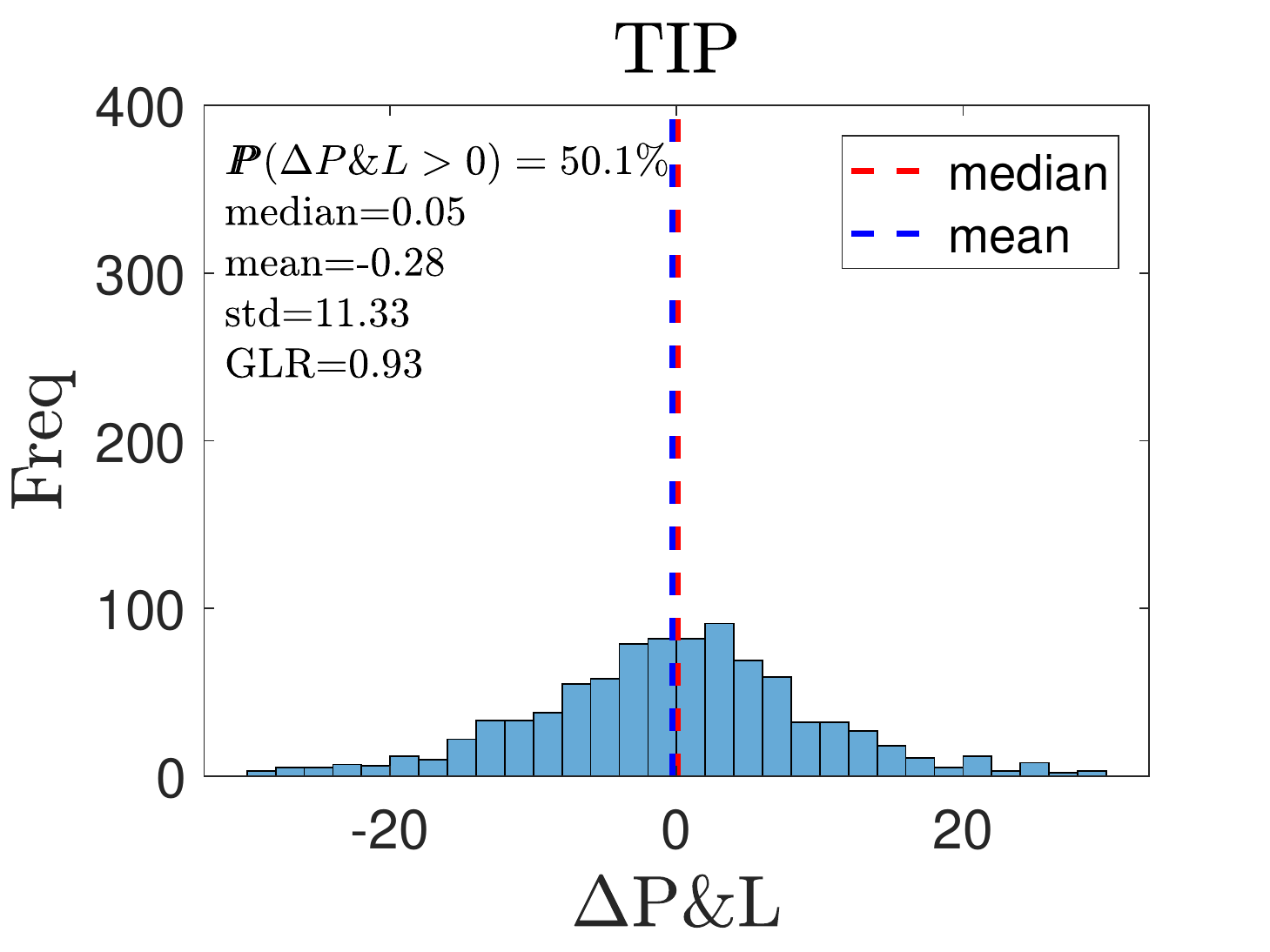}}
&
\vcenteredhbox{\includegraphics[width=0.32\textwidth]{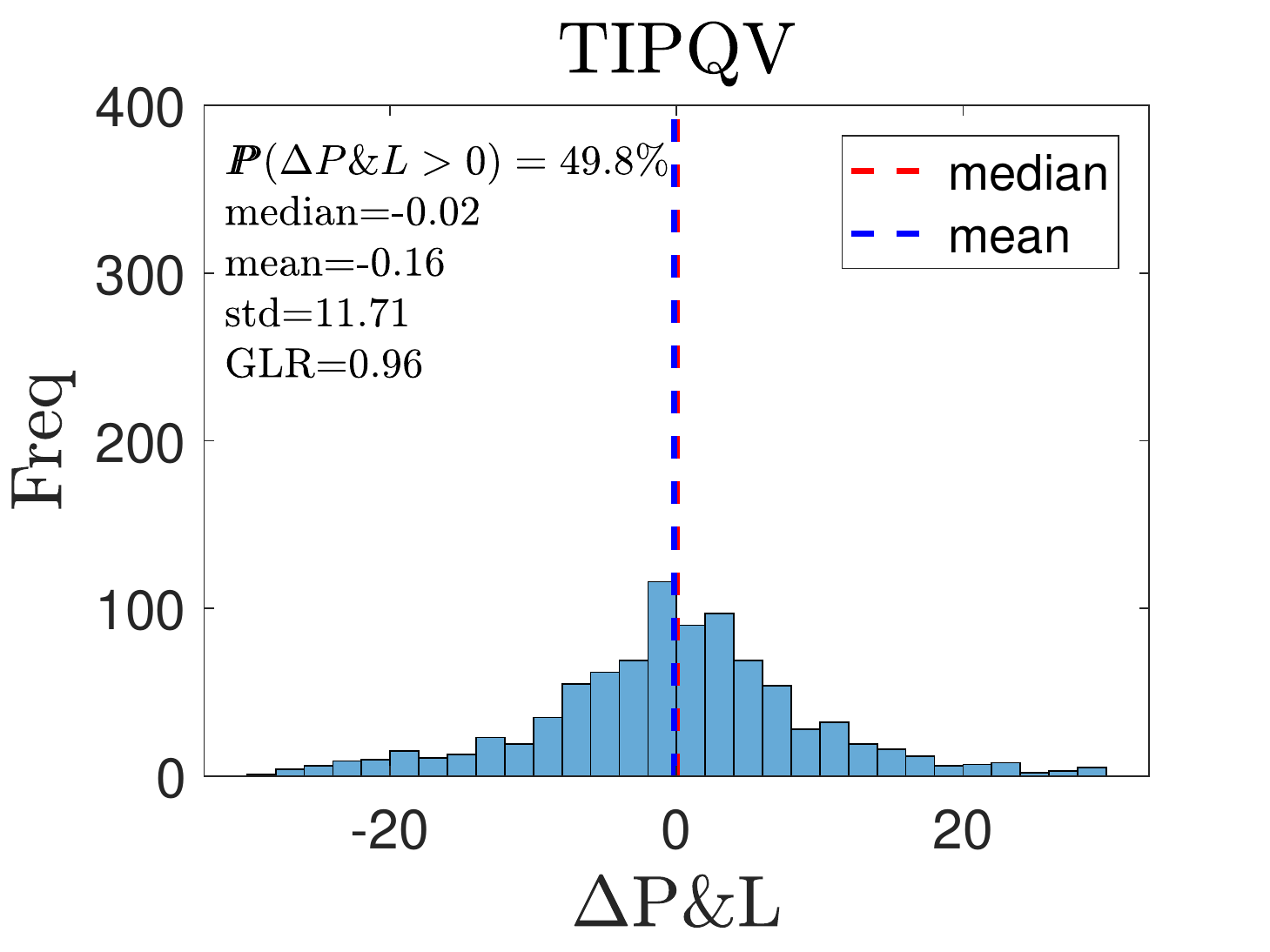}}
\\[1em]
FB &
\vcenteredhbox{\includegraphics[width=0.32\textwidth]{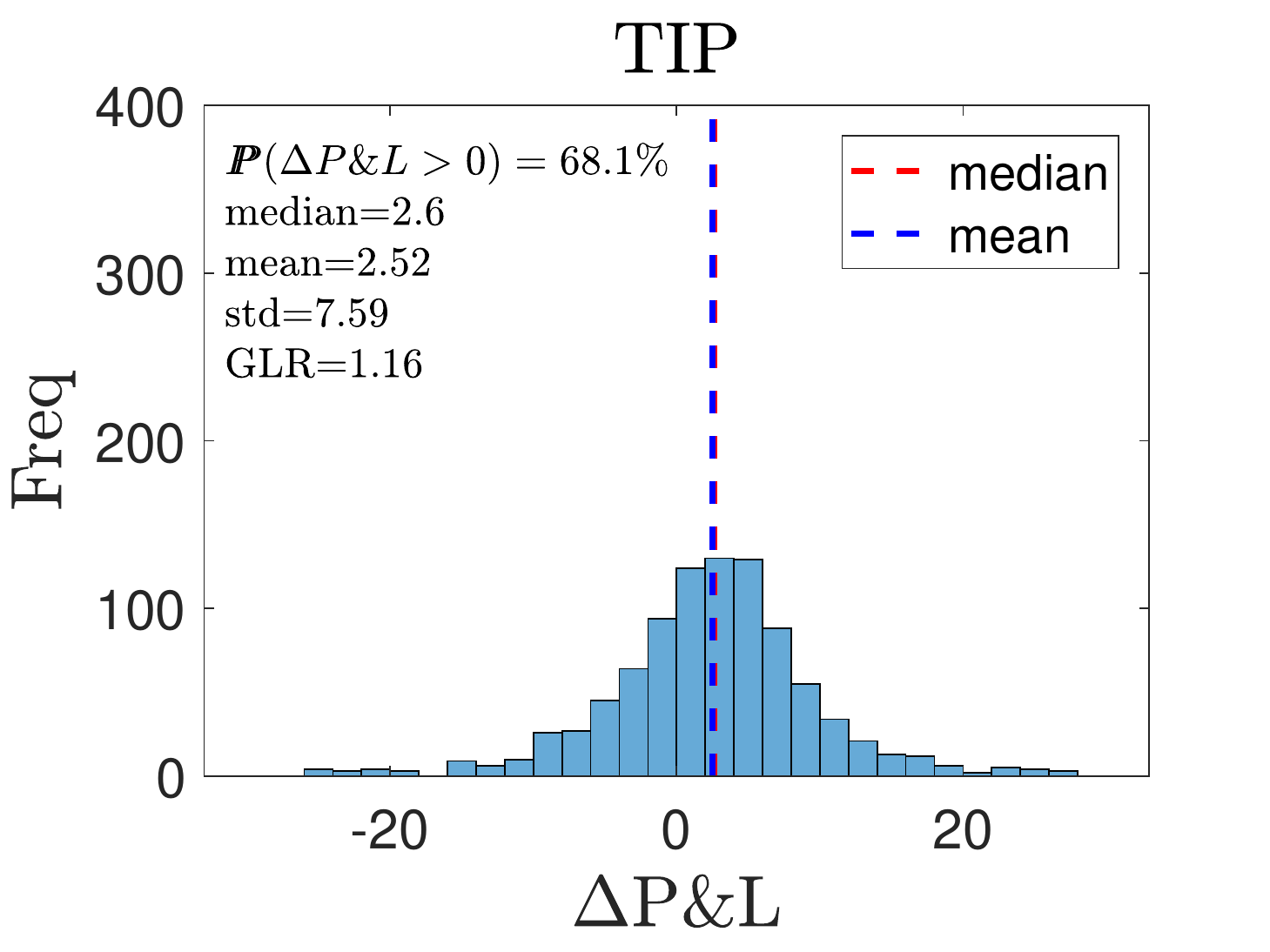}}
& \vcenteredhbox{\includegraphics[width=0.32\textwidth]{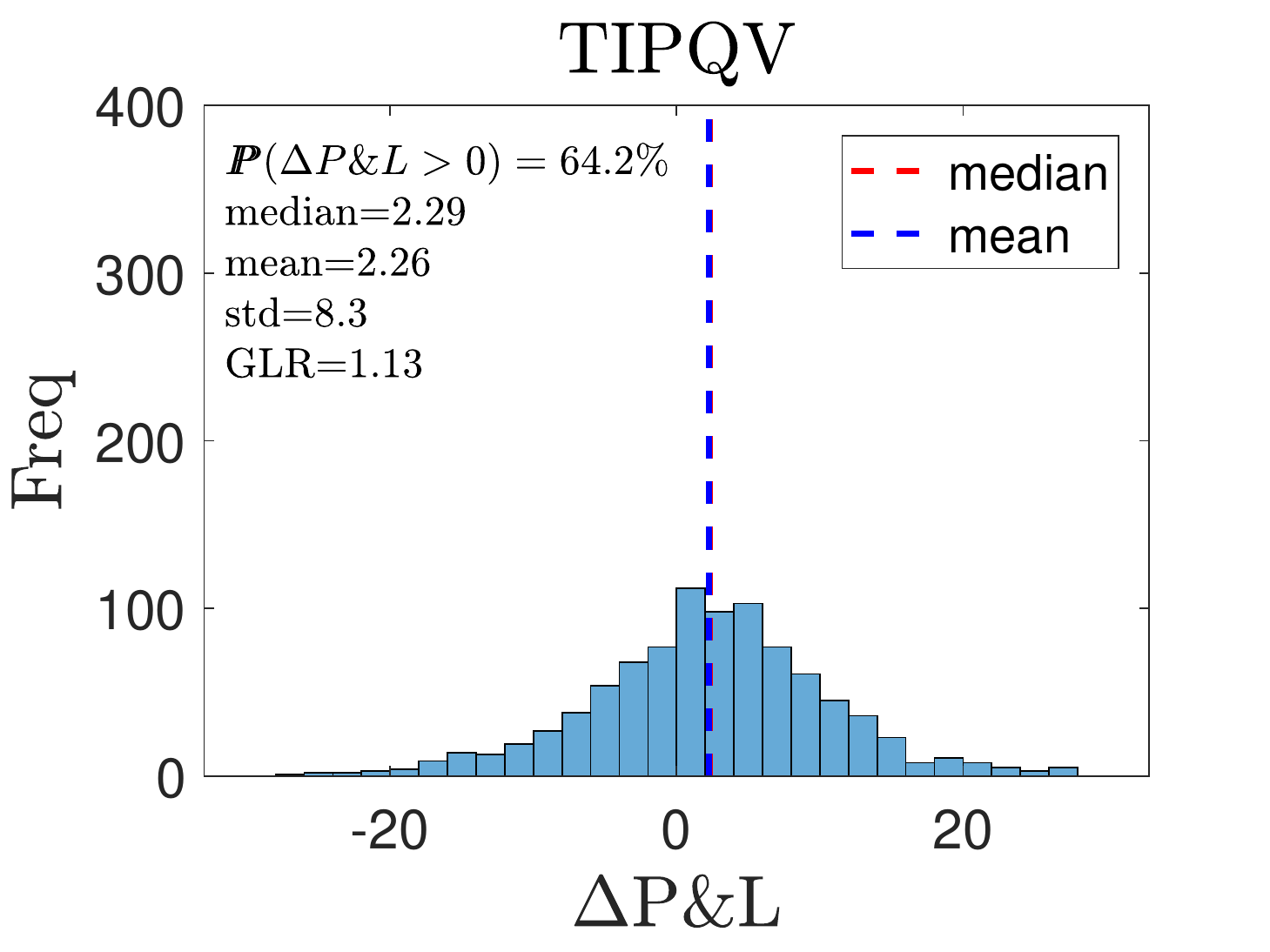}}
\\[1em]
GOOG &
\vcenteredhbox{\includegraphics[width=0.32\textwidth]{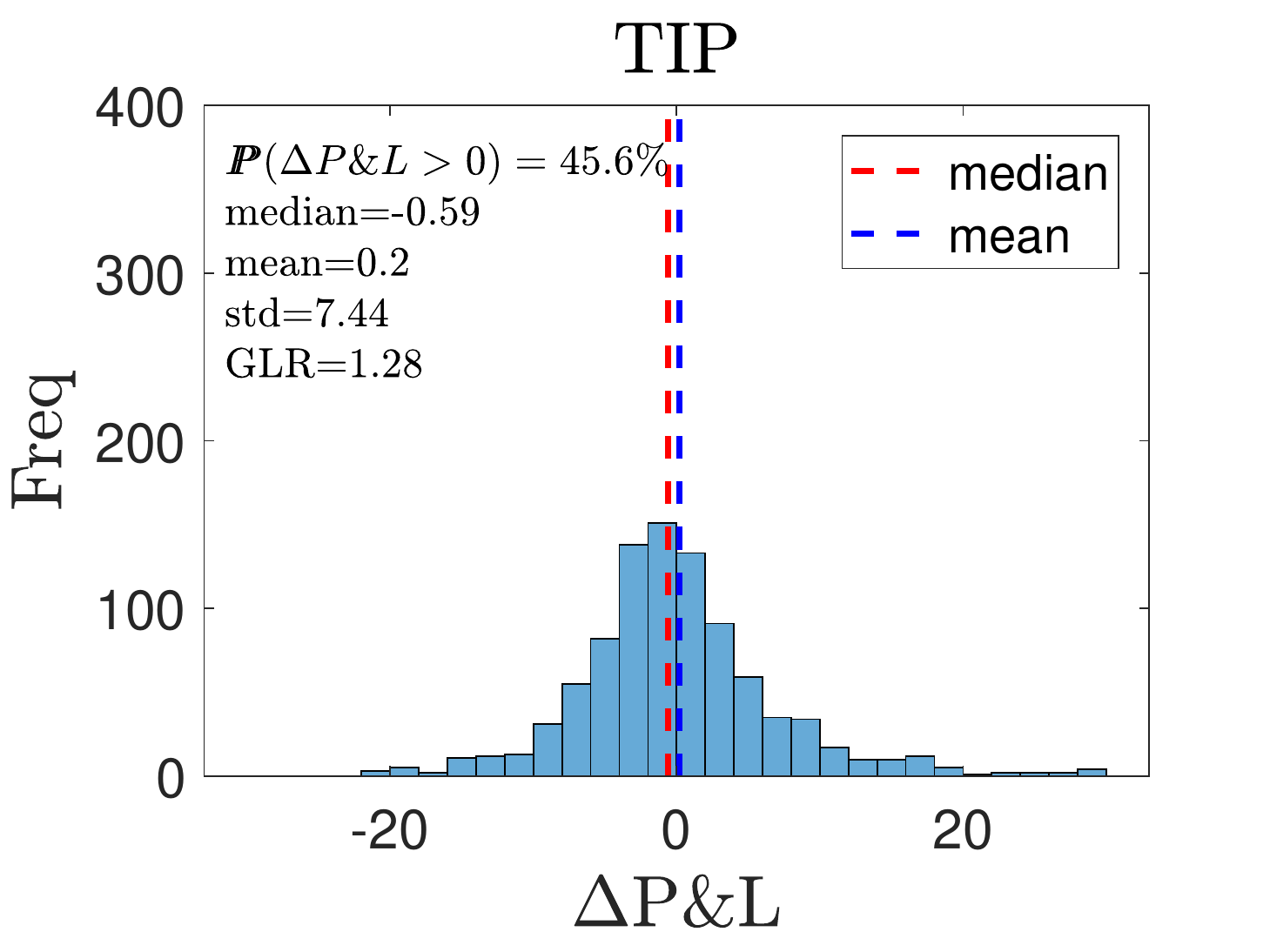}}
&
\vcenteredhbox{\includegraphics[width=0.32\textwidth]{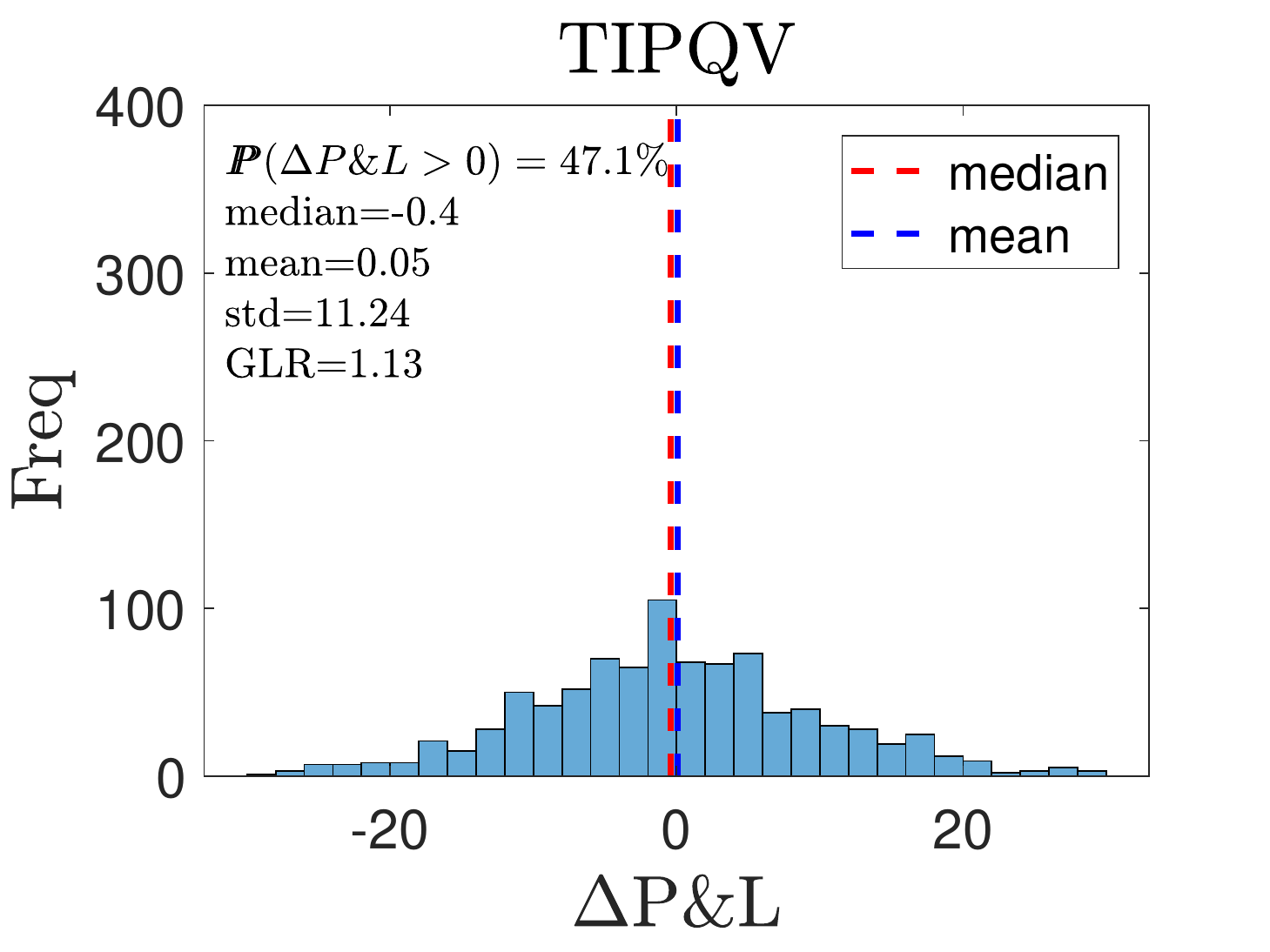}}
\\[1em]
INTC &
\vcenteredhbox{\includegraphics[width=0.32\textwidth]{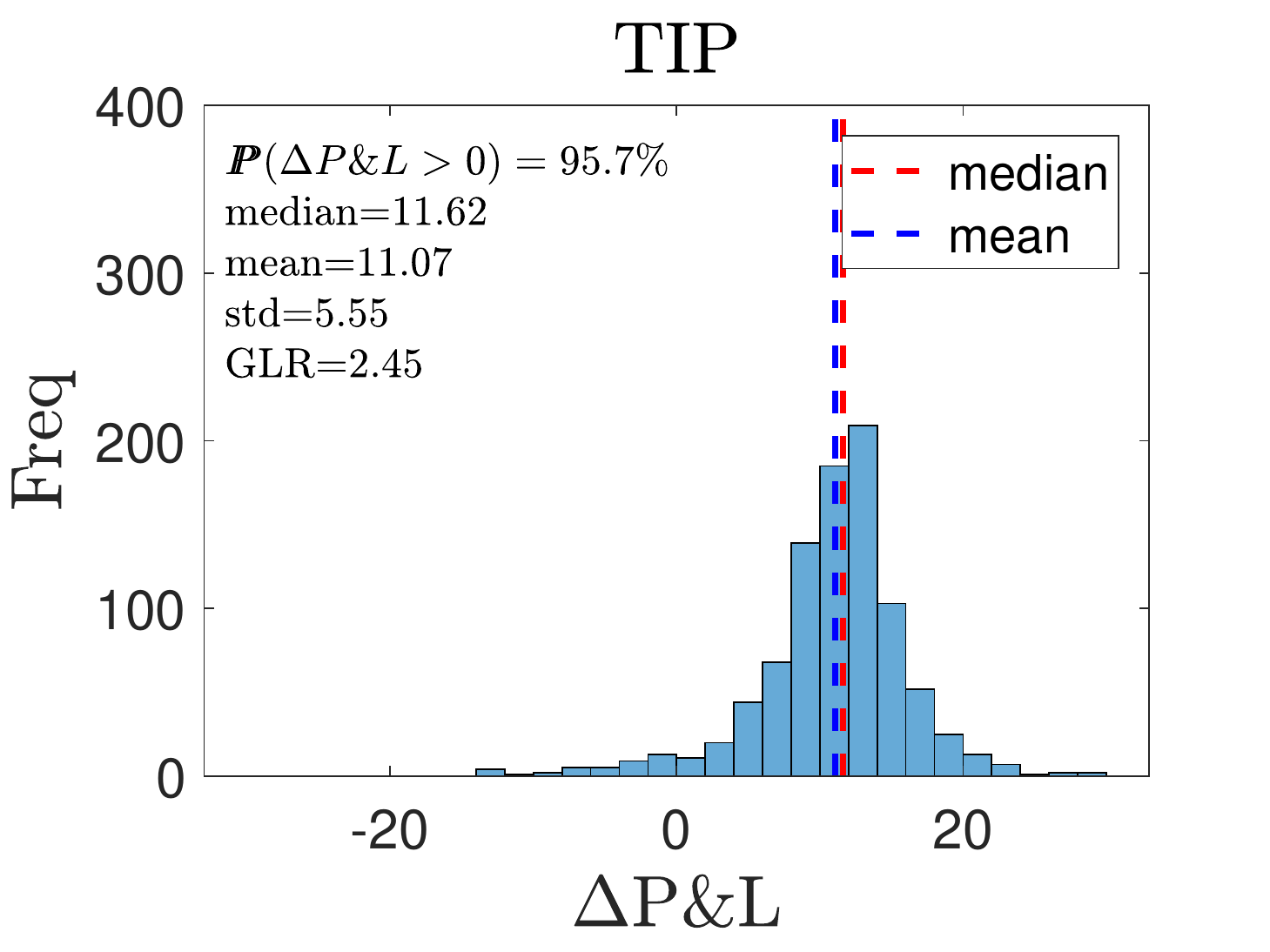}}
&
\vcenteredhbox{\includegraphics[width=0.32\textwidth]{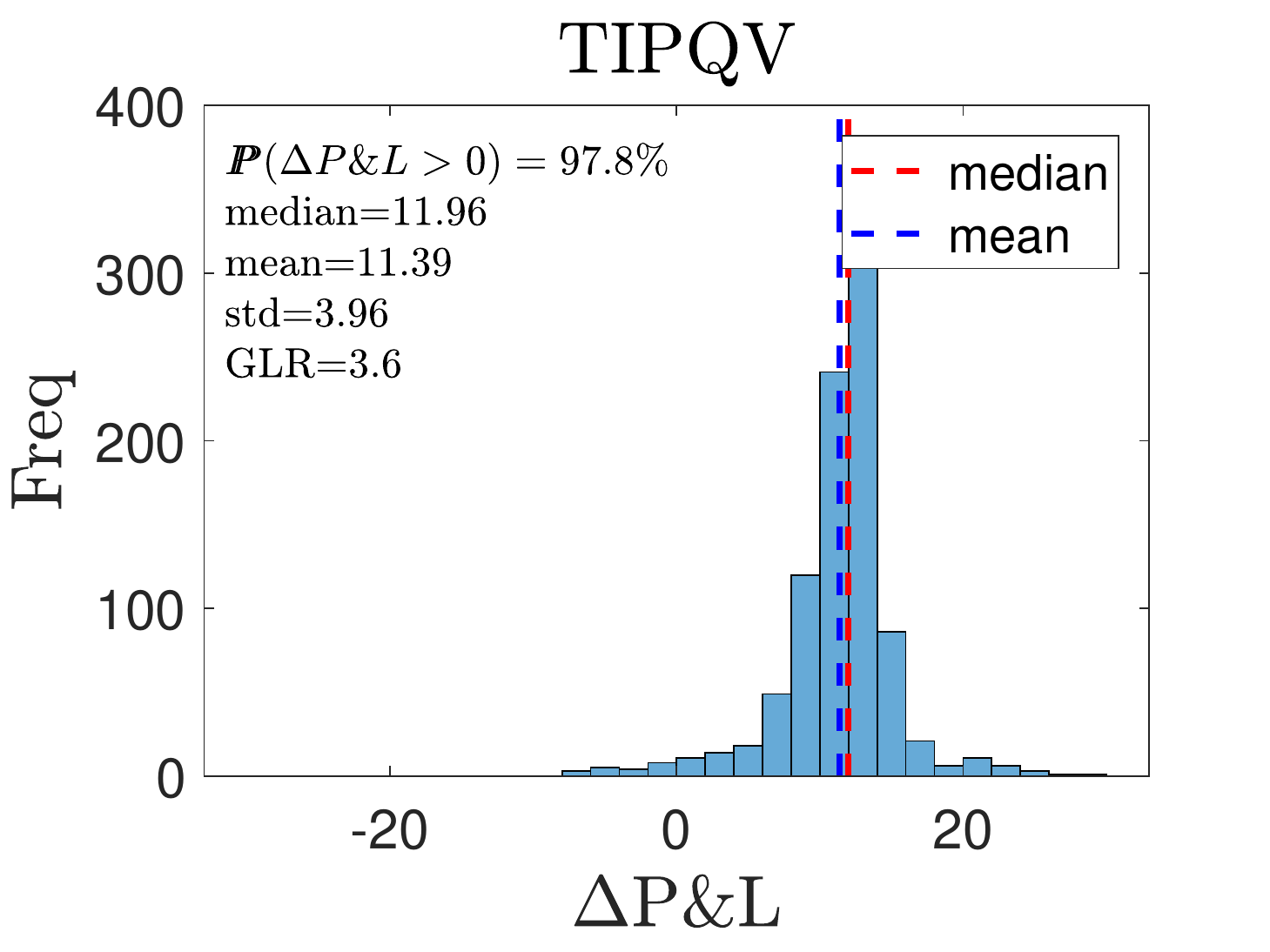}}
\end{tabular}
\end{center}
\caption{Distribution of the relative P\&L (basis points) with various features.
\label{fig:all-results-1}}
\end{figure}

\begin{figure}[h!]
\setlength{\tabcolsep}{10pt}
\renewcommand{\arraystretch}{1.5}
\begin{center}
\begin{tabular}{rcc}
MSFT
&
\vcenteredhbox{\includegraphics[width=0.32\textwidth]{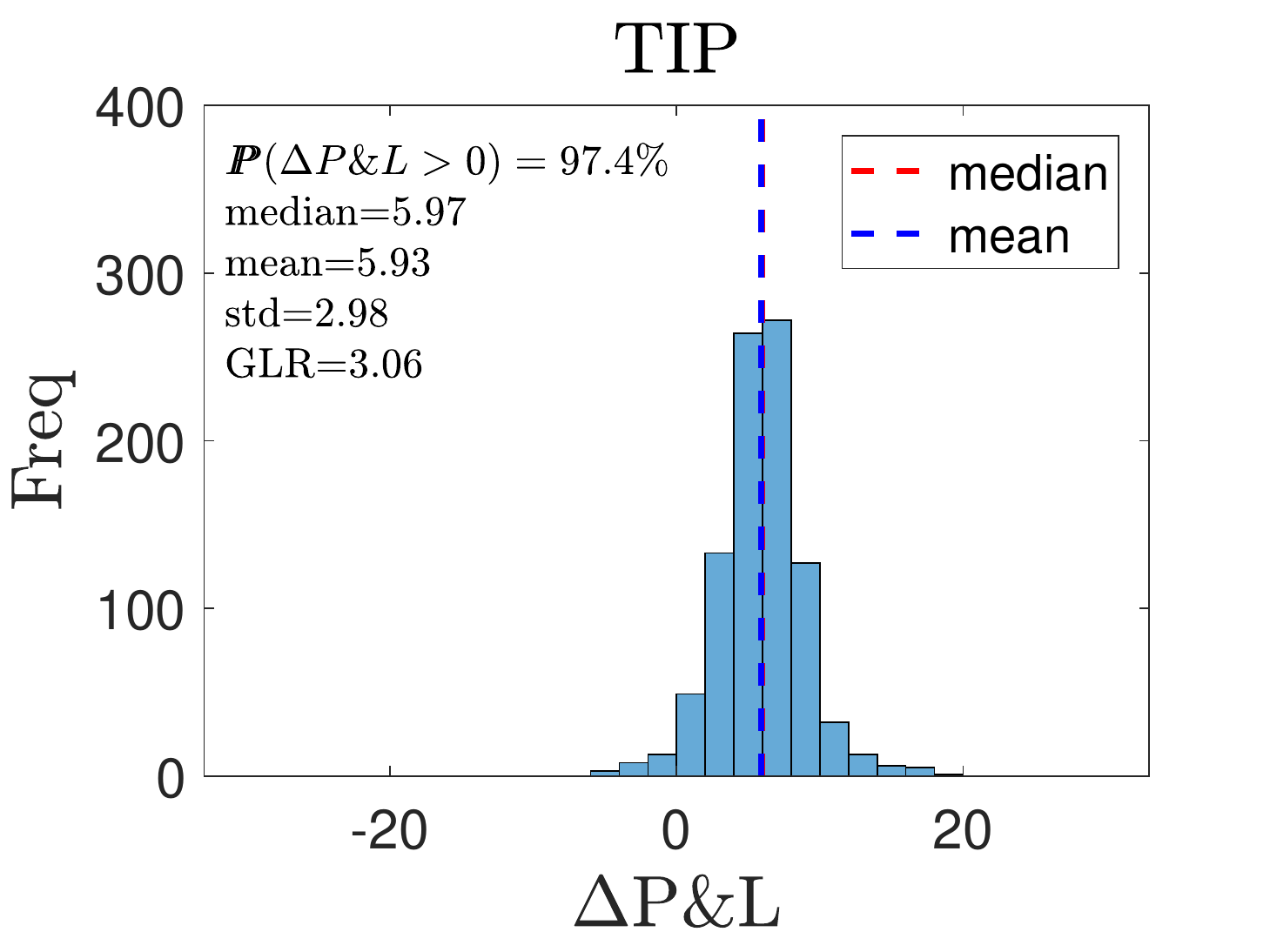}}
&
\vcenteredhbox{\includegraphics[width=0.32\textwidth]{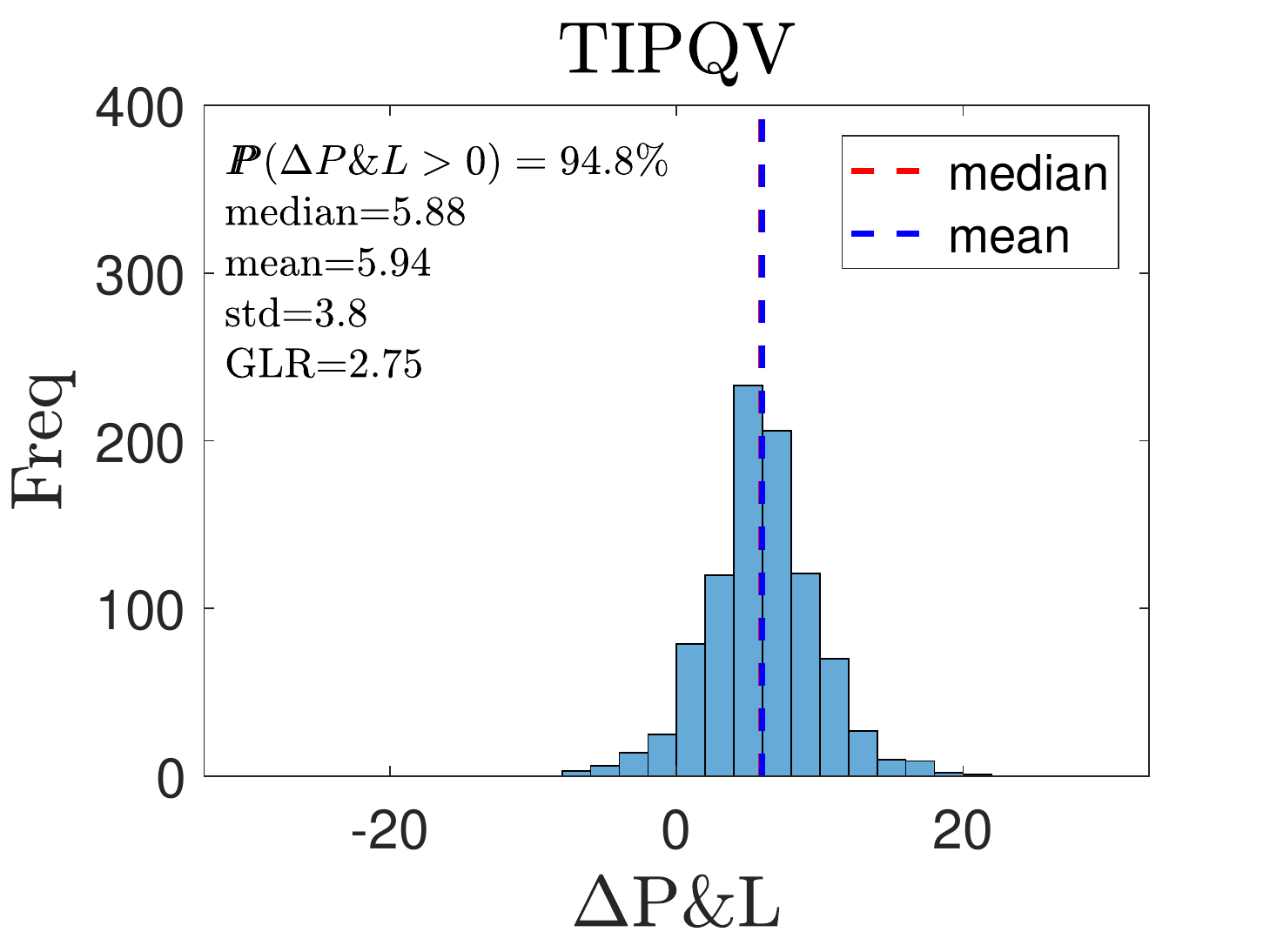}}
\\
NTAP
&
\vcenteredhbox{\includegraphics[width=0.32\textwidth]{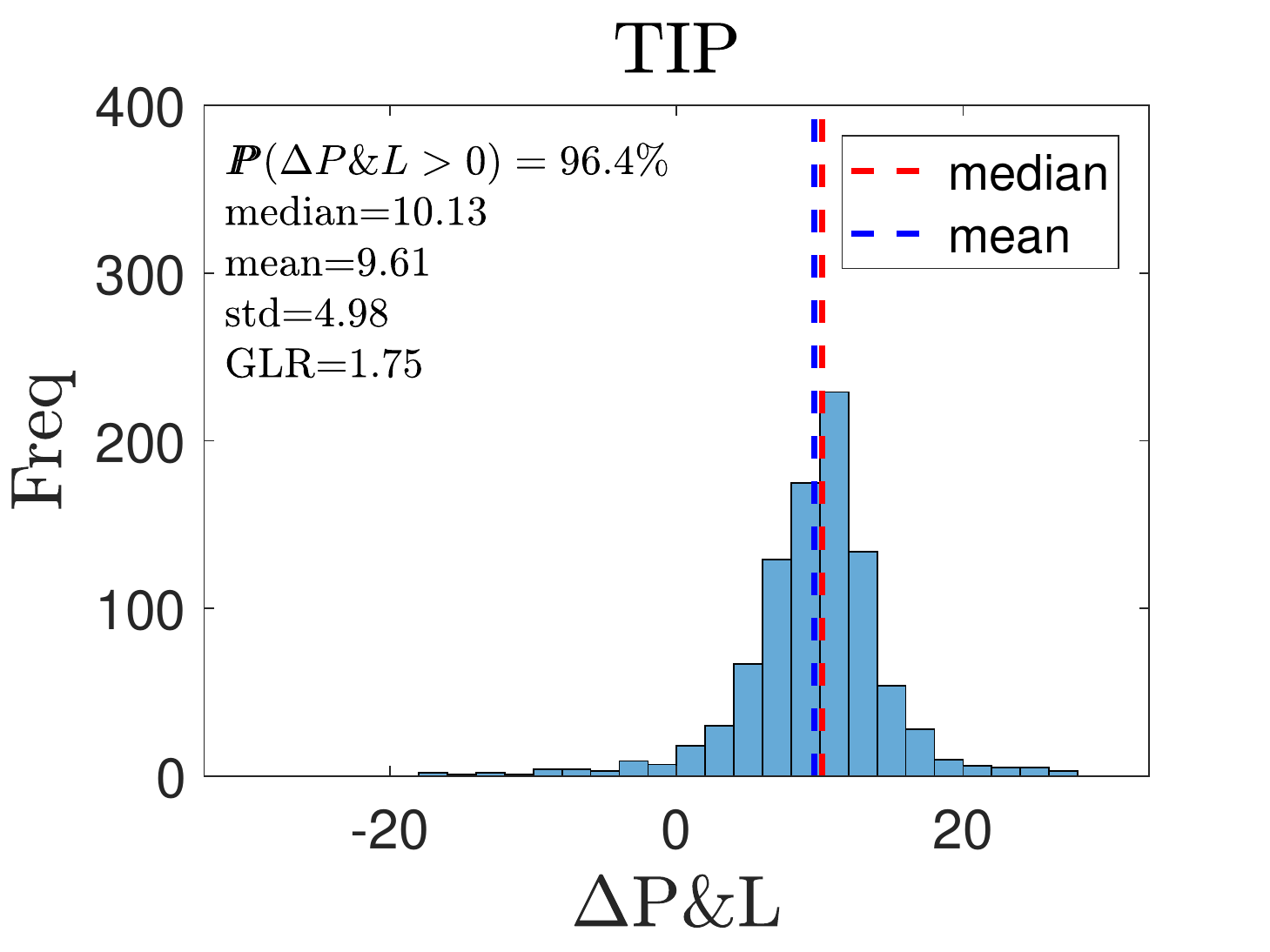}}
&
\vcenteredhbox{\includegraphics[width=0.32\textwidth]{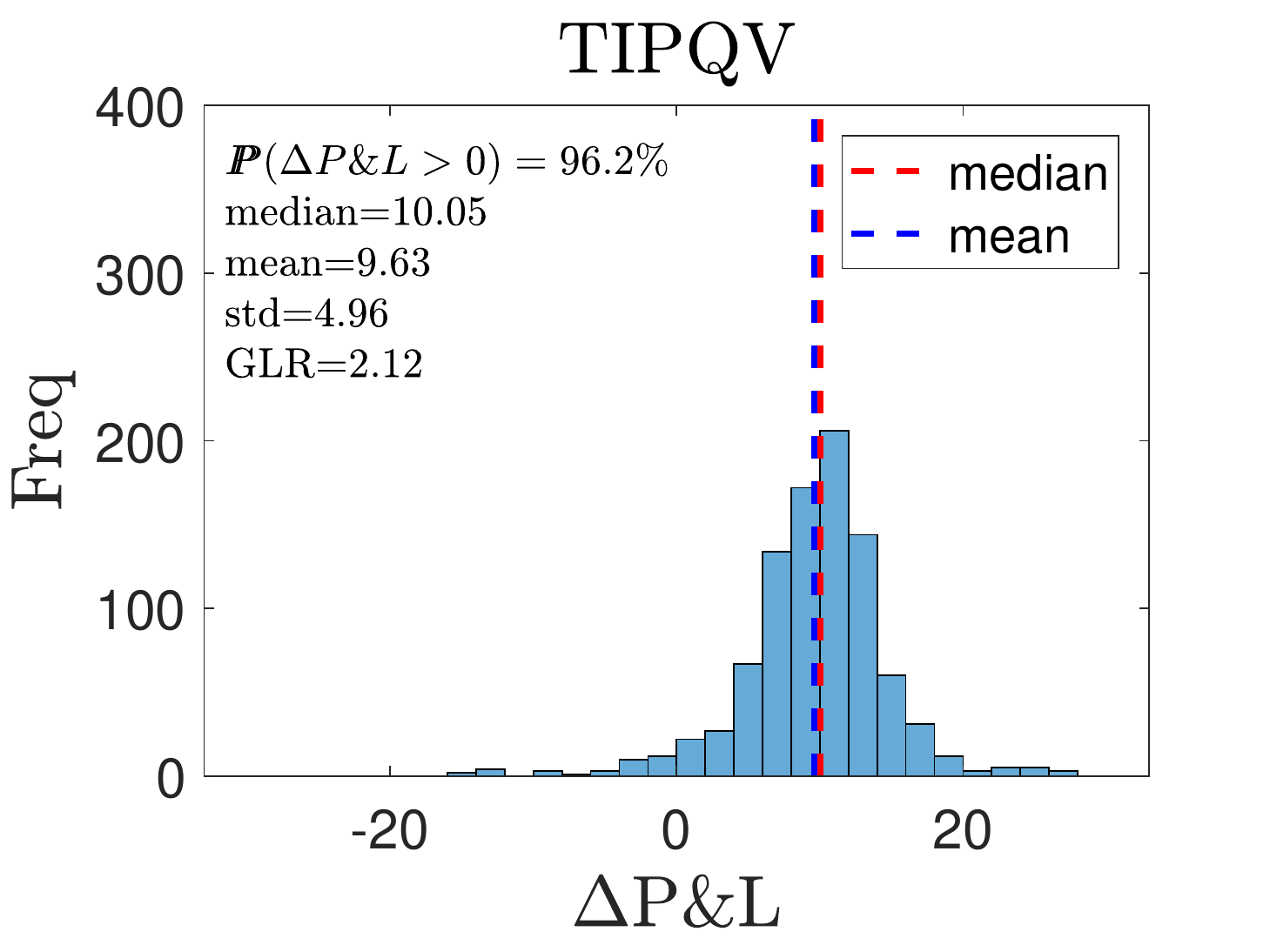}}
\\
SMH
&
\vcenteredhbox{\includegraphics[width=0.32\textwidth]{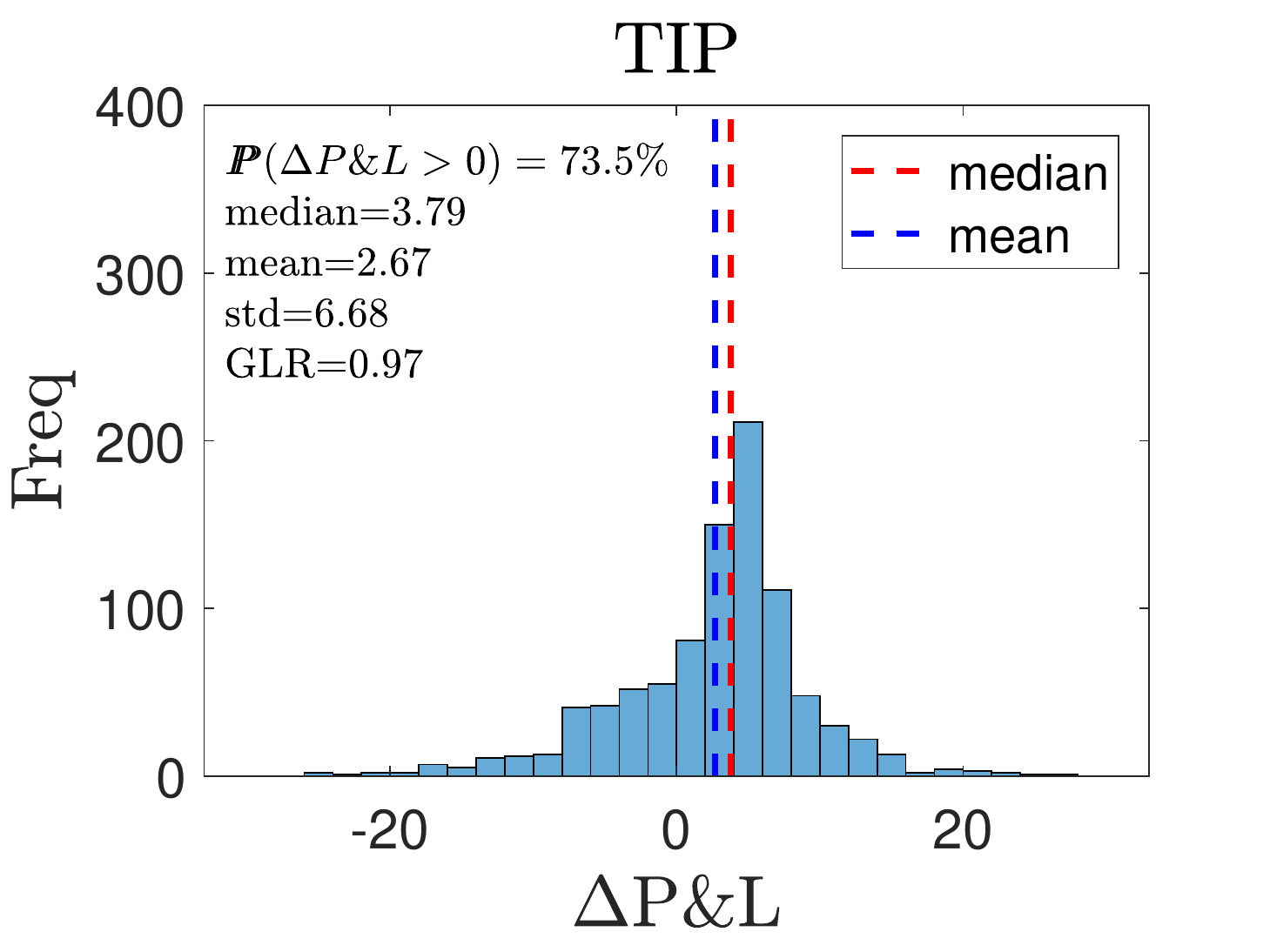}}
&
\vcenteredhbox{\includegraphics[width=0.32\textwidth]{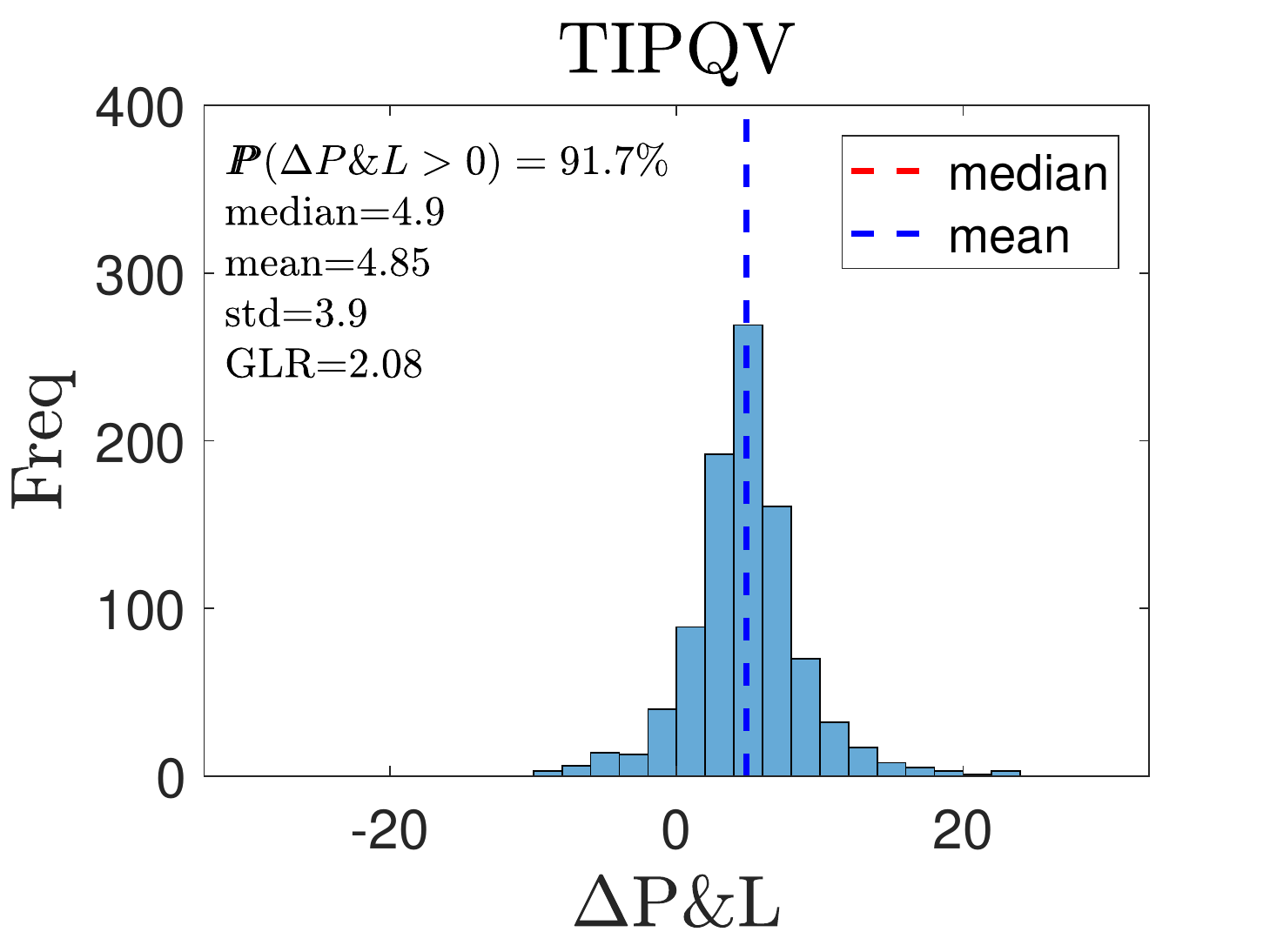}}
\\
VOD
&
\vcenteredhbox{\includegraphics[width=0.32\textwidth]{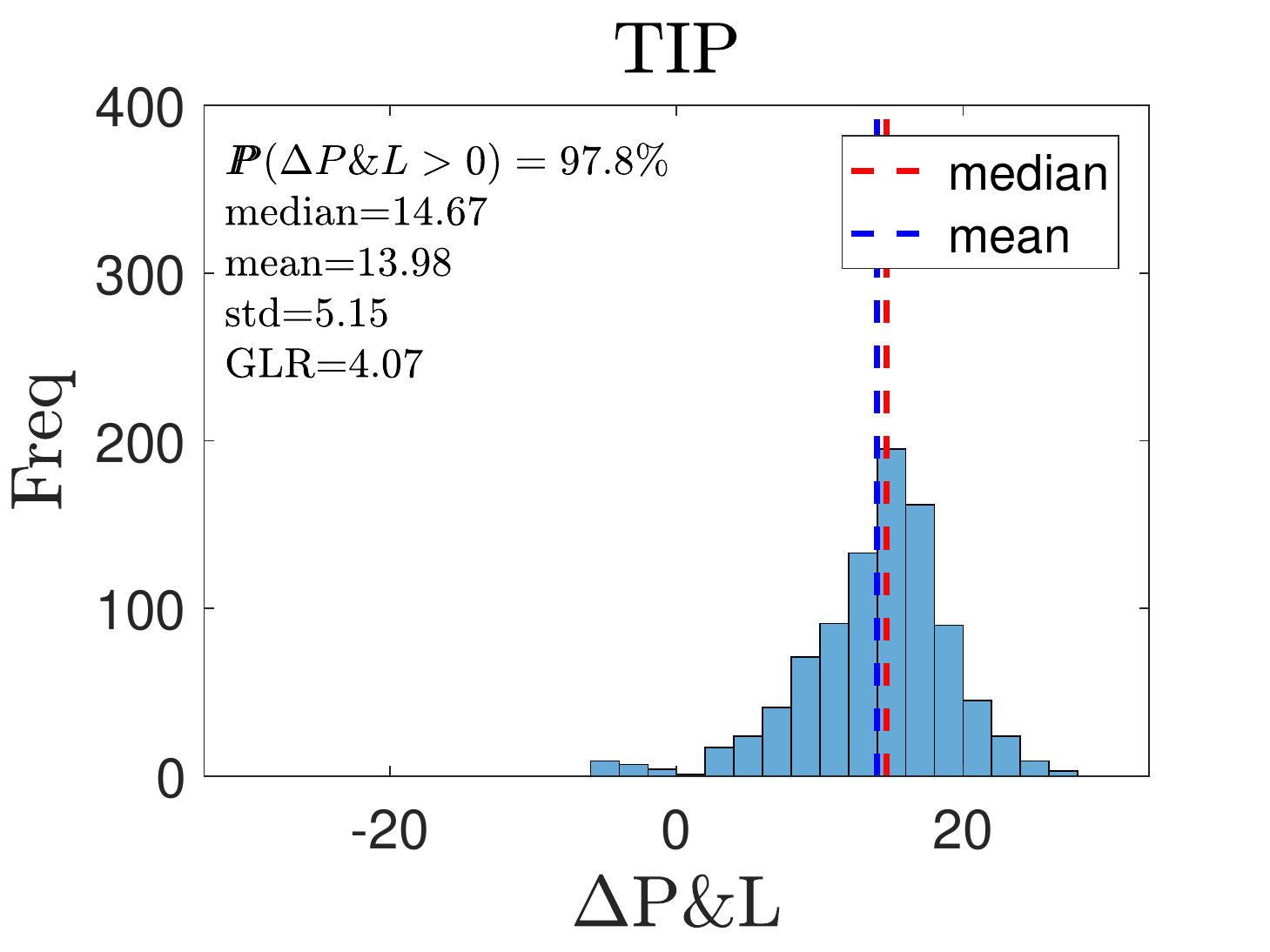}}
&
\vcenteredhbox{\includegraphics[width=0.32\textwidth]{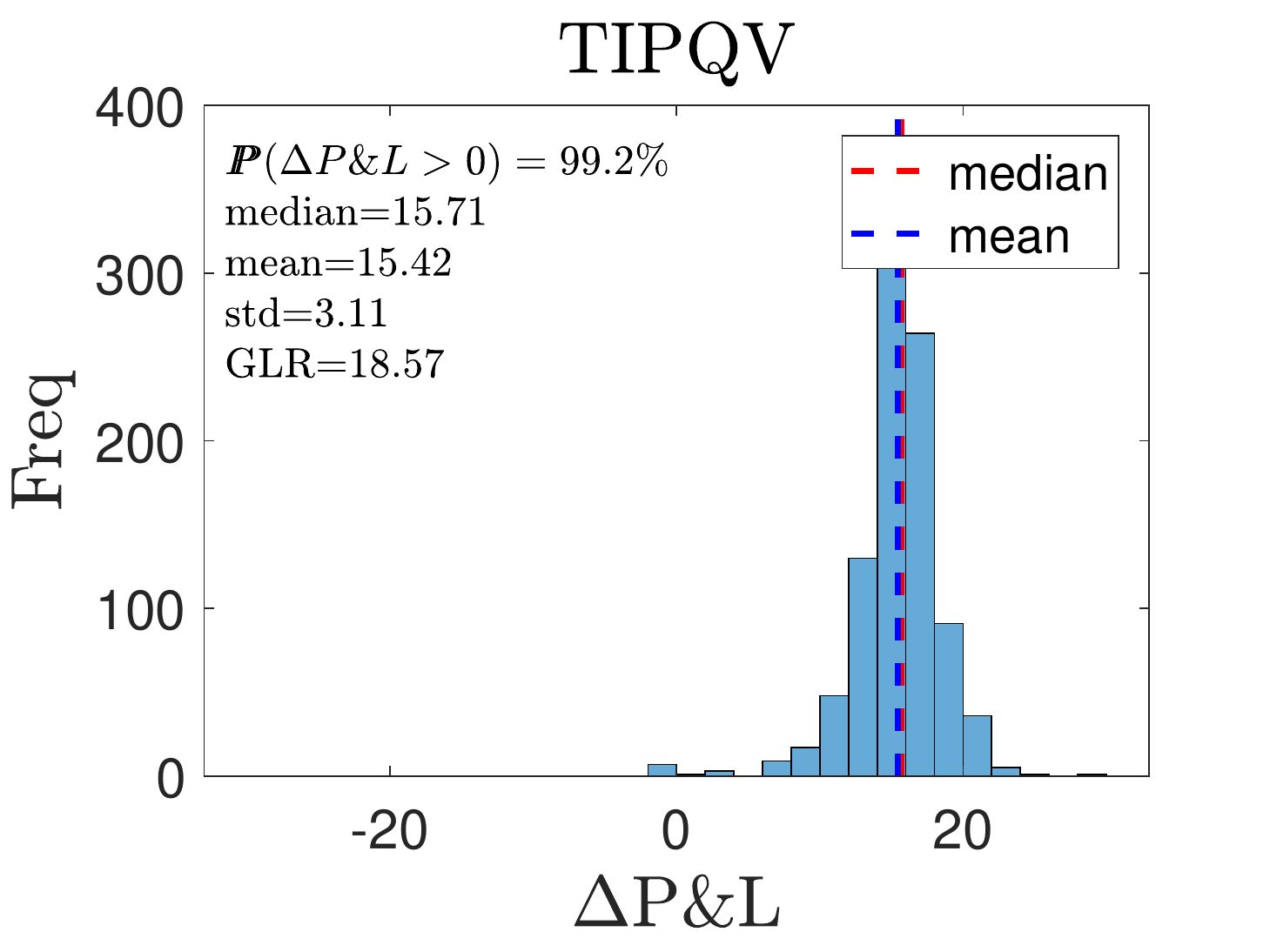}}
\end{tabular}
\end{center}
\caption{Distribution of the relative P\&L (basis points) with various features.
\label{fig:all-results-2}}
\end{figure}

The corresponding histograms of the relative P\&L are shown in Figures \ref{fig:all-results-1} and \ref{fig:all-results-2}.

\clearpage

\section{Conclusions}
\label{sec:conclusions}
In this paper we formulated the optimal execution problem as a reinforcement learning problem. We developed a deep reinforcement learning technique that requires a number of modifications to the double deep Q-learning approach to account for the hard constraints as well as the . The results show the approach  outperforms TWAP on seven out of the nine stocks and for the two under performing stocks, one is statistically insignificant. There are a number of directions still left open for investigation. Two obvious direction are to increase the number of assets analysed and to include a number of additional features, such as, price history and limit order book history. Another direction is to combine the analytical approaches, e.g., \cite{cartea2016incorporating} and \cite{casgrain2018}, or data-driven modeling as in \cite{sirignano2018deep}, together with deep Q-learning by using the analytical optimal trading strategy as the starting point for reinforcement learning.


\bibliographystyle{siamplain}
\bibliography{references}



\appendix
\section{Appendices}

\subsection{Transforming Inventory-Action Domain}
\label{sec:transform-q-x-domain}

For all inventory-action pairs $\{(q,x): (q,x) \in (0,q_0]^2, \; x\le q\}$, we perform the following non-linear transformation into the domain $[-1,1]^2$. First, shift and normalize
\begin{equation}
\hat{q} = \tfrac{q}{q_0} -1, \qquad
\hat{x} = \tfrac{x}{q_0}.
\end{equation}
Next, define the radial distance, angle, and ratio as
\begin{equation}
\begin{aligned}
r &= \sqrt{q^2 + x^2}, \\
\theta &= \tan^{-1}\left(-\frac{x}{q}\right), \qquad \text{and}\\
\zeta &= -\frac{x}{q},
\end{aligned}
\end{equation}
respectively. Transform the radial distance via
\begin{equation}
	\tilde{r} =
   \begin{cases}
      r \sqrt{(\zeta^2 + 1)\,(2\, \cos^2(\frac{\pi}{4} -\theta)) }, & \theta \leq \frac{\pi}{4} \\[0.5em]
      r \sqrt{ (\zeta^{-2} + 1)\,(2 \,\cos^2(\theta - \frac{\pi}{4}))   }, & \theta > \frac{\pi}{4}.
   \end{cases}
\end{equation}
Finally convert the polar coordinates into the domain $[-1,1]^2$ to produce the features
\begin{equation}
\begin{aligned}
\tilde{q} &= -\tilde{r}\,\cos(\theta)\,,\qquad \text{and} \qquad
\tilde{x} &= \tilde{r}\,\sin(\theta)\,.
\end{aligned}
\end{equation}
This produces the transformation of the inventory/action pairs as shown in Figure \ref{fig:trans}.

\end{document}